\documentclass{jfm}
\usepackage{graphicx}
\usepackage{bm}
\usepackage{subfig}
\usepackage{float}
\usepackage{subfloat}
\usepackage{epstopdf, epsfig}
\usepackage{amsmath}
\usepackage{amssymb}
\usepackage{tikz}
\usetikzlibrary{arrows}
\usepackage{tikz-3dplot}
\usepackage{hyperref}
\usepackage{pgf,pgfplots,pgfplotstable}
\usetikzlibrary{decorations.markings}
\usetikzlibrary{patterns}
\usetikzlibrary{intersections,shapes.arrows}
\usetikzlibrary{arrows.meta, bending, decorations.markings}
\usetikzlibrary{spy}
\usepackage{mathrsfs}
\usetikzlibrary{angles,quotes}
\usetikzlibrary{arrows}
\usepackage{siunitx}
\usetikzlibrary{calc,decorations.pathmorphing}
\usetikzlibrary{plotmarks}

\usetikzlibrary{decorations.markings}
\tikzset{
    set arrow inside/.code={\pgfqkeys{/tikz/arrow inside}{#1}},
    set arrow inside={end/.initial=>, opt/.initial=},
    /pgf/decoration/Mark/.style={
        mark/.expanded=at position #1 with
        {
            \noexpand\arrow[\pgfkeysvalueof{/tikz/arrow inside/opt}]{\pgfkeysvalueof{/tikz/arrow inside/end}}
        }
    },
    arrow inside/.style 2 args={
        set arrow inside={#1},
        postaction={
            decorate,decoration={
                markings,Mark/.list={#2}
            }
        }
    },
}

\tikzset
  {every pin/.style={pin edge={<-}}
  ,>=stealth
  ,flow/.style=
    {decoration=
      {markings
      ,mark=at position #1 with {\arrow{>}}
      }
    ,postaction={decorate}
    }
  ,flow/.default=0.5
  }

\newcommand\inlaycaption[1]{{\sffamily\scriptsize#1}}
\newcommand\newinlay[4][0.18]%
  {\renewcommand\inlayscale{#1}%
   \newsavebox#2%
   \savebox#2%
     {\begin{tabular}{@{}c@{}}
        #4\\[-1ex]
        \inlaycaption{#3}\\[-1ex]
      \end{tabular}%
     }%
  }

\newcommand{\be}[1]{ \begin{equation} \label{#1}}
\newcommand{\ee}{\end{equation}}

\newcommand{\bes}[1]{ \begin{equation} \label{#1}\begin{array}{rl}}
\newcommand{\ees}{\end{array}\end{equation}}

\newcommand{\bu}{ {\bf u} }
\newcommand{\ddp}[2]{\frac{\partial #1}{\partial #2}}

\graphicspath{{./}{./FIGURES_NACA_JUIN2025}{./FIGURES_NACA_JUIN2025/FIGURES/}}

\shorttitle{Transition in strongly convective flows }
\shortauthor{D. Fabre}

\title{Is the transition to unsteadiness in the wake of slender bodies an artefact of boundary conditions ?  }

\author{
 David Fabre
 }

\affiliation{
Institut de M\'ecanique des fluides de Toulouse (IMFT), University of Toulouse, France. }

\begin{document}
\maketitle
\begin{abstract}

This work considers the transition to unsteadiness in the wake of 2D slender bodies, and questions the relevance of the generally accepted scenario involving a region of absolute instability within the near wake. The case of  a thin plate at zero incidence is first considered. Despite the absence of absolute instability region, global stability analysis  reveals the existence of numerous unstable eigenvalues organized along a characteristic "arc-branch" whose properties significantly depends upon the size of the numerical domain. These arc-branch modes are explained as resulting from a non-local pressure perturbation spuriously generated at the outlet of the domain due to the no-stress boundary condition, which then triggers the shedding of vortical structures at the trailing edge of the plate. The case of NACA0012 wing profiles at small incidences is then considered. Global stability analysis reveals that both the non-local spurious feedback mechanism and the classical local feedback mechanism are active. Trying to suppress the spurious feedback by enlarging the size of the numerical domain is shown to be inefficient. On the other hand, filtering methods suppressing the exponential spatial growth of perturbations, with either a sponge or a complex mapping, are found to be efficient.
Thanks to these ideas, the critical Reynolds number and Strouhal number at onset can eventually be computed and are mapped for incidences in the range $ \alpha \in [0^o - 5 ^o]$. It is postulated that the non-local feedback mechanism evidenced here could be at play in other strongly convective flows.

\end{abstract}

\section{Introduction}

The transition from a steady state to an unsteady time-periodic state is ubiquitously observed in spatially developing open flows, including wakes, jets, shear layers, and is usually the first transition in a process leading to a fully turbulent state. Among wake problems, the generic case of 2D blunt bodies, leading to the well-known Von-Karman vortex alley, has served as a guideline for the conceptualization of hydrodynamic stability theory. The transition is well predicted using global stability analysis of the preexisting steady base flow, and following a well-accepted view (see the review papers of \cite{HuerreMonkewitz1990}, \cite{Chomaz2005} and references herein), it is well explained in terms of local stability concepts as resulting from the existence of a region of absolute instability of sufficient extend, roughly identified with the recirculation region. 
This explanation was initially theorized for the case of blunt bodies, the most studied case being the 2D cylinder where the bifurcation occurs at a critical Reynolds number  $Re_c \approx 46.7$.

Periodic vortex shedding is also observed for the case of slender bodies, and the scenario presented above is generally expected to remain valid, although the transition occurs at much larger values of the Reynolds number. Among slender bodies, NACA profiles are often considered as prototypes of 2D wing profiles and have been considered in a number of study. Recent studies have drawn a comprehensive picture of both the cambered NACA4412 profile \citep{nastro2023global} and the symmetric NACA0012 profile \citep{victoria2022stability,gupta2023two}. However there studies focused on configurations with an angle of attack larger than $5^o$, corresponding to situations where a strong recirculation region exists due to detachment of the boundary layer on the extrados. The picture for small incidences is less clear. \citep{gupta2023two} have hypothesized a power law relating the critical Reynolds number to the angle of attack with form $\alpha \approx Re_c^{-0.65}$, indicating that the critical Reynolds should tend to infinity in the limit of zero incidence. This is not confirmed by \cite{sabino2022aeroelastic} who investigated the zero-incidence case with DNS. This study reported the emergence of periodic shedding in the range $Re \approx [7000-8000]$, but this emergence was observed to be gradual,  first appearing very far away from the wing. Moreover, a strong dependance with respect to the mesh resolution and domain dimension was observed, which did not allow to determine a precise value for the critical Reynolds number.
Note also that, unlike the transition to unsteadiness in a 2D flow discussed here, the subsequent transition leading to three-dimensional flow (which occurs at $Re \approx 30\,000$ for zero incidence) seems much easier to deal with, and was investigated using Floquet analysis by both \cite{sabino2022aeroelastic} an  \citep{gupta2023two}.

Aside for wing profiles, the even more fundamental case of a thin plate with zero incidence also made the object of a few studies leading to apparently conflicting results. The WKBJ analysis of \cite{woodley1997global}, later corrected by \cite{taylor1999note}, shows that the wake does not contain regions of absolute instability, thus normally excluding the existence of unstable global modes. \cite{pralits2009global} Investigated the problem with global stability approach. Their paper only presents results relevant to finite-thickness profiles modeled by a Falker-Skan boundary layer profile at the inlet of the domain, but according to personal discussions with these authors, they also observed unstable global eigenmodes for the zero-thickness profile which they were not able to explain.

Aside from wake flows, there is another class of open flows with provided a fruitful playground for global and local stability approaches, namely jets.
According to local approach, a pure jet in incompressible homogenous flow is only convectively unstable, hence self-sustained oscillation is not expected. However, a number of additional ingredients lead to emergence of global instability, including temperature inhomogeneity and buoyancy \citep{Monkewitz1988}, swirl \citep{Nichols2011}, coupling with an acoustic resonator \citep{sierra2022acoustic}, or impact upon a plate \citep{ausin2023mode}. In such cases, a number of studies have reported numerical difficulties linked to the effect of domain truncation. For instance, \cite{fabre2019acoustic} showed that the calculation of the impedance of a jet, which is directly linked to its instability properties, is strongly affected by the outlet boundary conditions, leading them to propose a complex mapping method allowing to get rid of this spurious effect. In global stability analysis of inhomogeneous jets, 
\cite{coenen2017global} and \cite{chakravarthy2018global} have identified that domain truncation indeed leads to the existence of spurious modes called "arc-branch" modes as they occur in global spectra as a family of eigenvalues localized along a characteristic arc-like curve. Recognizing that such arc-branches are a generic feature in global stability of open flows in a truncated domain, \cite{lesshafft2018artificial} gave a detailed description of such spurious modes. More specifically, analysis of the pressure component of the eigenmode showed that it contains two parts, the first associated to the convectively amplified local mode, and the second being nonlocal and resembling eigenfunctions of the Poisson equation.  However, this study considered homogeneous Neumann conditions for the pressure on all boundaries, which is not the most commonly adopted choice, and did not address the possibility interplay of this nonlocal pressure and the convective instability of the jet. Nevertheless, \cite{lesshafft2018artificial} proposed an absorbing layer as an effective means to filter spurious modes due to boundary conditions. However, the simplified cases considered in his study contained only spurious arc-branch modes. In more physically meaningful cases, spurious arc-branch modes can be expected to coexist with physically relevant modes, and it is crucial to design a method allowing to filter the first family without affecting the second.

In the present paper, the problem of the transition in the wake of slender bodies will be revisited using global stability. 
After presentation of the methodology in \S2, the case of a flat plate of zero-thickness thickness will first be considered in \S 3, showing that arclength modes are present in this case and are dominating the spectrum.
In \S 4, the nature of these arc-branch modes will be explained through the interplay of convective instability of the wake and a non-local pressure component generated at the outlet boundary as a artefact of the no-stress condition. Subsequently,  \S 5 turns back to the NACA0012 profile at zero incidence, and shows that the difficulties encountered in this case are due to a competition between spurious arc-branch modes an a physically relevant eigenmode linked to absolute instability in the recirculation region. Is is eventually shown that the arc-branch modes can be filtered using specifically designed filtering techniques, yielding access to the prediction of transition in the idealized  case of a slender body "in an infinite 2D domain". Eventually, in \S 6, this method is used to map the marginal stability curve of a NACA0012 profile at small incidences, filling a gap in the literature.

\section{Methodology}
\subsection{Governing equations}
Throughout this paper we will consider a flow field governed by the incompressible Navier-Stokes equation :
\be{NS}
\frac{\partial \bu}{\partial t} + ( \nabla \bu ) \cdot \bu = - \nabla p + Re^{-1} \Delta \bu ; \quad \nabla \cdot \bu = 0.
\ee
Where $\bu = [u,v]$ is the velocity and $p$ the pressure. Here the single physical control parameter is the Reynolds number $Re$ based on the inlet velocity and  the chord of the wing, both normalized as 1.  The problem is formulated in a truncated rectangular domain of dimensions $x\in [-X_i; L_x]$,  $y\in [-Y_L; Y_L]$, the origin of the frame being taken at the trailing edge of the wing. At the inlet and lateral boundaries, a uniform velocity  
$[u,v] =[1,0]$ is applied. 
On the other hand, at the outlet boundary, a no-stress condition is imposed, and the dimension $L_x$ in the downstream dimension will be shown to have a significant effect of the results. 

Before moving on, it is worth pointing out two important properties of the set of equations and boundary conditions used here. 
First, in the incompressible case, the pressure is governed by an elliptic equation; indeed, taking the divergence of the Navier-Stokes equations leads to:
\be{Poisson}
\Delta p  = - [\nabla \bu] :  [\nabla \bu]^T.
\ee 
This equation, being nonlocal, allows an instantaneous communication between all parts of the domain. Secondly, it can be remarked that, 
in the range of Reynolds number of interest here, the normal stress 
grossly correspond to the pressure ; so the no-stress condition imposed at the outlet of the domain actually amounts to imposing a constant pressure.
These two points will prove to be essential in the nonlocal feedback model to be presented in \S 3.
 
 \subsection{Global stability analysis}

Global stability analysis starts classically by the expansion
\be{GlobalAnsatz}
[{\bf u},p] = [{\bf u}_b(x,y),p_b(x,y)] + \epsilon  [\tilde{{\bf u}}(x,y),\tilde{p}(x,y)] e^{- i \omega t}
\ee
where  $[{\bf u}_b(x,y),p_b(x,y)]$ is the base flow, namely the steady solution of the Navier-Stokes equations,
and where 
  $\tilde{\bf{q}} = [\tilde{\bf u}(x,y),\tilde{p}(x,y)]$ is an eigenmode associated to a complex frequency $\omega = \omega_r + i \omega_i$, where $\omega_r$ is the oscillation rate and $\omega_i$ is he amplification rate. 
Solutions of the linearised Navier-Stokes equations which can be set in the form of an eigenvalue problem, written in block-matrix form as follows:

\be{EigenvalueProblem}
 {\bf A} \tilde{\bf{q}} = - i \omega  {\bf B} \tilde{\bf{q}}, \quad 
\mbox{with} \quad A = 
\begin{bmatrix}
- \nabla  {\bf u_b} \cdot ( \bullet ) - \nabla ( \bullet ) \cdot {\bf u}_b   + Re^{-1} \Delta (\bullet)  & - \nabla ( \bullet ) \\ 
\nabla \cdot (  \bullet ) & 0 
\end{bmatrix}, \quad 
B = 
\begin{bmatrix}
{\bf 1}  & 0 \\  0 & 0  
\end{bmatrix}.
 \ee

 \subsection{Local stability analysis}
 
In situations where the base-flow is parallel (or can be locally approximated as parallel), one can classically use a so-called {\em local stability approach}, 
which consists of considering eigenmodes with modal dependence in the axial direction $x$, parametrized by an axial wavenumber $k$. Namely, we start from the ansatz \ref{GlobalAnsatz} with the following additional hypotheses :
\be{LocaAnsatz}
[{\bf u}_b,p_b] = [U(y) {\bf e}_x, 0] ; \qquad  [\tilde{{\bf u}}(x,y),\tilde{p}(x,y)] e^{- i \omega t} =  [\hat{\bf u}(y),\hat{p}(y)] e^{ik x} e^{-i \omega t}.
\ee

In this framework, the linear eigenvalue problem \ref{EigenvalueProblem} can be classically reduced to a single equation governing the transverse velocity component  $\hat{u}_y(y)$, called the Orr-Sommerfeld equation : 
\be{OS}
{\cal L}_{OS} ( \hat{u}_y ) \equiv (k U- \omega) \left( \partial_y^2 - k^2\right)  \hat{u}_y  + k U'' \hat{u}_y + i Re^{-1} \left( \partial_y^2 - k^2\right)^2  \hat{u}_y 
 = 0
 \ee
 This problem has the advantage of being a 1-D problem in terms of the function $\hat{\bf u}_y(y)$, and is thus much more easily solved than the global stability problem presented above.
 
Two viewpoints are generally adopted to solve this equation, and both will be useful in the sequel. The first point of view is the {\em temporal} analysis which consists of solving the problem for $\omega$ (complex) as function of $k$ (real). 
The second viewpoint, which is more suited to describe amplification in the downstream direction in convective problems, is the {\em spatial analysis}, namely,  the OS eigenproblem is solved for $k=k_r+i k_i$ (complex) as function of the frequency $\omega$ (assumed real). In this framework, $-k_i$ is the spatial amplification rate, and $\omega/k_r$ is the phase velocity.



 \subsection{Numerical methods}
 
In the sequel, all computations are done using a finite-difference method thanks to the FreeFem++ software. Time-stepping (DNS) computations are done using a fully implicit, second order (BDF2)  time discretization, as described in \cite{ausin2023mode}. The global stability calculations are done following the classical approach described in \cite{Fabre2019}.
Both the time-stepper and the eigenvalue solver are parallelized thanks to PETSc/SLEPc, and the computations are typically ran over 30 processors.  
Note that mesh adaptation is heavily used to achieve a satisfying spatial convergence ; here adaptation is done using as target fields both the base-flow, the leading eigenmode and the associated adjoint eigenmode. As demonstrated in \cite{Fabre2019}, this
approach is efficient to achieve results independent from the mesh (except of course from effects from the boundary conditions) for global stability computations.  For DNS computations additional constraints are introduced in the mesh adaptation process to achieve a sufficient resolution in the wake region where nonlinear dynamics occur. All computations and post-processing are monitored thanks to the StabFem toolbox, and sample codes demonstrating all steps of the computations can be found on the website of the projet.

\section{Numerical results for a thin plate in a truncated domain}
\subsection{Time-stepping results} 

\begin{figure}
\includegraphics[width=.48\linewidth]{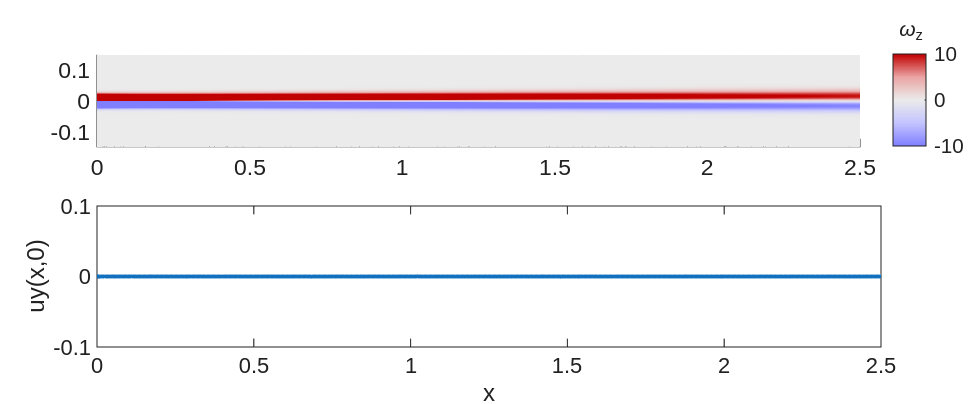}
\includegraphics[width=.48\linewidth]{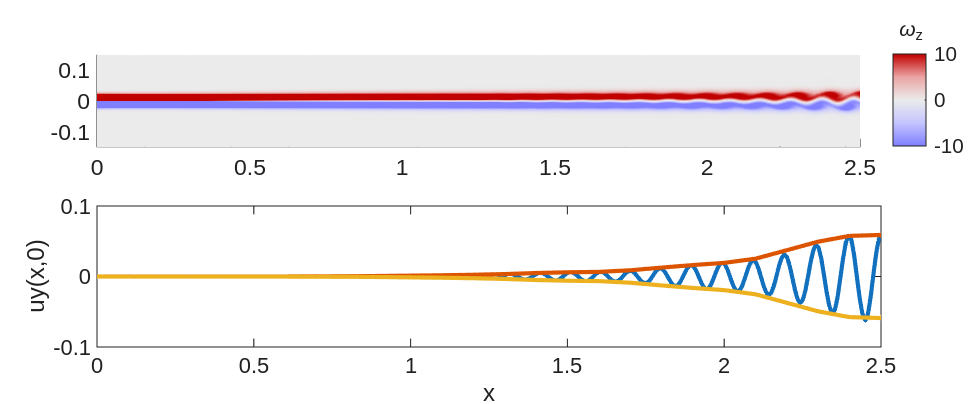}
$$
\includegraphics[width=.89\linewidth]{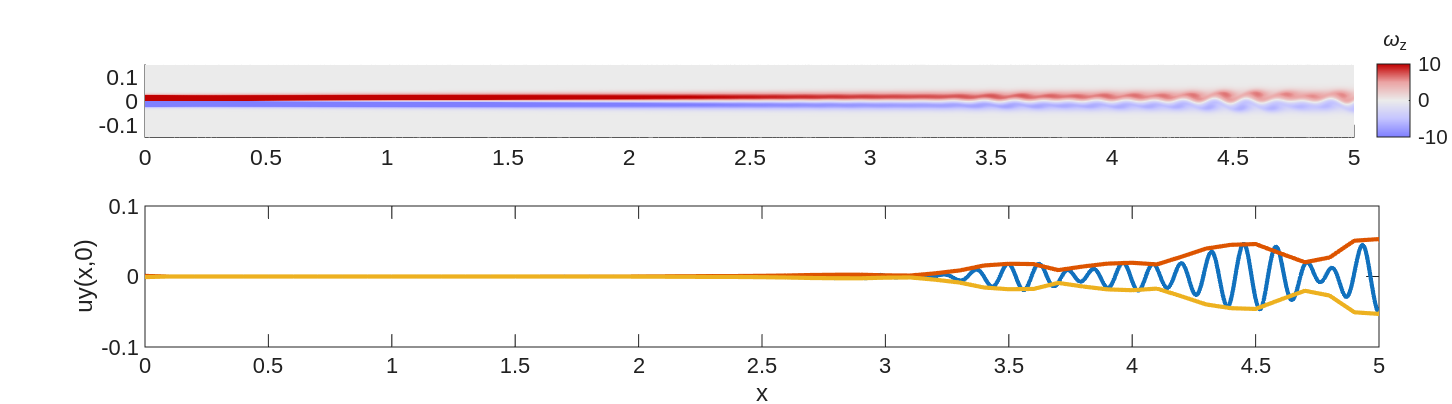}
$$
$$
\includegraphics[width=.89\linewidth]{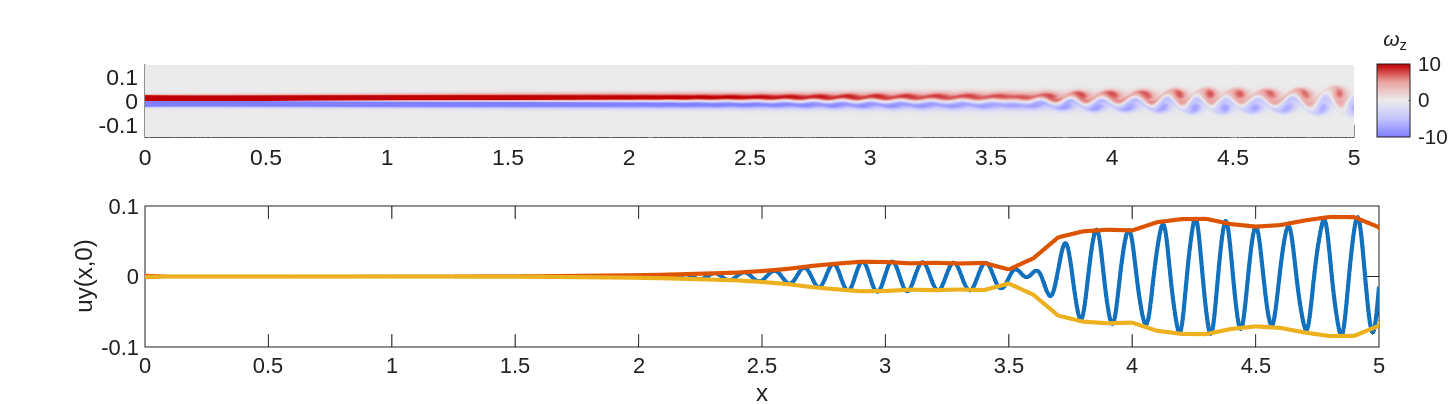}
$$
 \caption{Sample DNS results for the flow around a thin plate in a truncated domain. $(a)$ : $Re = 27\,500$, $L_x=2.5$; $(b)$ : $Re = 30\,000$, $L_x=2.5$; $(c)$ : $Re = 27\,500$, $L_x=5$; $(a)$ : $Re = 30\,000$, $L_x=5$. Upper plots $(a1,b1,...)$ are color maps of the vorticity; Lower plots  $(a2,b2,...)$ are transverse velocity $u_y$ along the axis, displaying the last DNS snapshot and the envelope of extreme values reached during the 10 previous oscillation periods.  }
 \label{fig:DNS_Plate}
 \end{figure}

The dynamics (later to be identified as spurious) observed in the wake of a thin plate will first be described through time-stepping simulations (DNS) of the incompressible equation in a truncated domain. Figure \ref{fig:DNS_Plate} displays DNS snapshots obtained for for $Re= 27\, 500$ and $30\,000$, 
and for two values of the downstream domain length, namely $L_x = 2.5$ and $5$, in units of the plate length. 
Note that these dimensions may look rater short ; however in the range of Reynolds number investigated the thickness of the boundary layer at the trailing edge is $\delta_0  \approx Re^{-1/2} = 0.006$, so these dimensions respectively correspond to $L_x= 400 \delta_0$ and $800 \delta_0$, hence the domain is actually quite long compared to the physically relevant length scales. The other dimensions of the domain are taken as very large, namely $L_{in} = Y_L = 20$, to get rid of any possible confinement effect and concentrate on the effect of $L_x$.

As can be seen in the figure, for the shortest domain $L_x=2.5$, the flow is found to converge towards a steady, symmetric state for $Re=27\,500$, (plot \ref{fig:DNS_Plate}$a)$ indicating stability of this base state. On the other hand, for $Re = 30 \, 000$, (plot \ref{fig:DNS_Plate}$b$) the flow is unsteady and eventually converges towards a periodic state, signaling the occurrence of a Hopf bifurcation, following the classical scenario for bluff bodies presented in the introduction. A different situation is observed for the longest domain; namely, unsteadiness is observed both for both values of $Re$.
Moreover, examination of the numerical solutions shows that contrary to expectation, the flow is not seen to transition to a strictly periodic state. Instead, a modulated oscillation is observed. The simulations were run on a rather long time, $t_{max} \approx 50$ corresponding to 10 convection 
length from the trailing edge to the outlet of the domain, and it was not possible to assert if the modulation is a very long transient towards a limit cycle 
or a chaotic attractor.  Nevertheless, this chaotic-like behavior is closer to the one expected for a strongly convective flow acting as a noise amplifier than the one of a self-sustained oscillator, following the terminology of \cite{HuerreMonkewitz1990}. However, in a noise amplifier, in absence of external forcing the perturbations eventually vanish. This is not the case here, indicating the existence of a feedback mechanism, which at this stage remains to be elucidated.

\begin{figure}
\includegraphics[width=.49\linewidth]{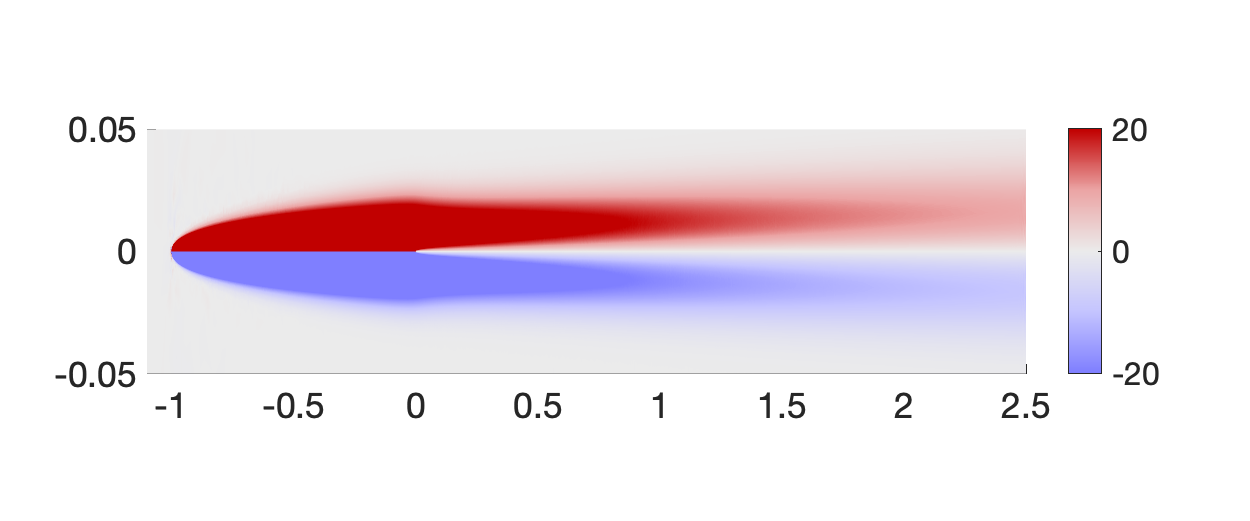}
\includegraphics[width=.49\linewidth]{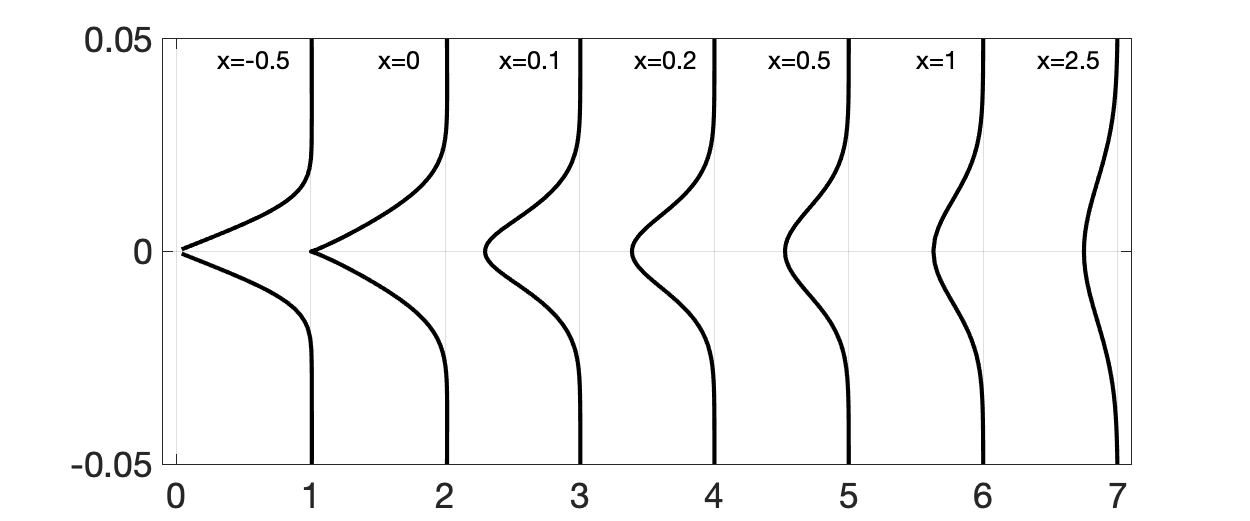}
\caption{Base flow around a flat plate for $Re = 27\, 500$. $(a)$ vorticity field (note the stretching in the $y$-axis) ; $(b)$ axial velocity profiles at several downstream locations. }
\label{fig:PlateBaseFlow}
\end{figure}

To complete this section on nonlinear simulation results,  figure \ref{fig:PlateBaseFlow} details the structure of the steady {\em base flow} obtained for $Re = 27 \, 500$. The left plot show a color plot of the vorticity field, as already displayed in figure \ref{fig:DNS_Plate}$(a)$ but using unequal axes to better represent the very thin shear-flow. Plot \ref{fig:PlateBaseFlow}$(b)$ displays profiles of the axial velocity $u_{b,x}$ as function of the transverse coordinate $y$ for several values of $x$. In the range $-1<x<0$, corresponding to locations along the plate, the velocity profile is very close to the Blasius boundary layer solution. After the plate ($x>0$), the velocity profile rapidly transitions towards a wake profile which can be fitted quite accurately with a Gaussian profile, with the form $U(y) \approx 1 - R \,e^{-y^2/\delta^2} $ where the velocity deficit $R$ and the thickness $\delta$ are functions of $x$.

Overall, the figure shows that the flow is locally very close to parallel. This property will allow, in a latter stage, to use local stability analysis to predict the growth of perturbations in the streamwise direction. Detailed results will be presented in a latter section (\S 3.3), but one can already note that the base flow does not possess any recirculation region (as the axial velocity is everywhere positive), hence does not contains any region of absolute instability.

\subsection{Global stability analysis}

\begin{figure}
\begin{tabular}{cc}
\includegraphics[width=.48\linewidth]{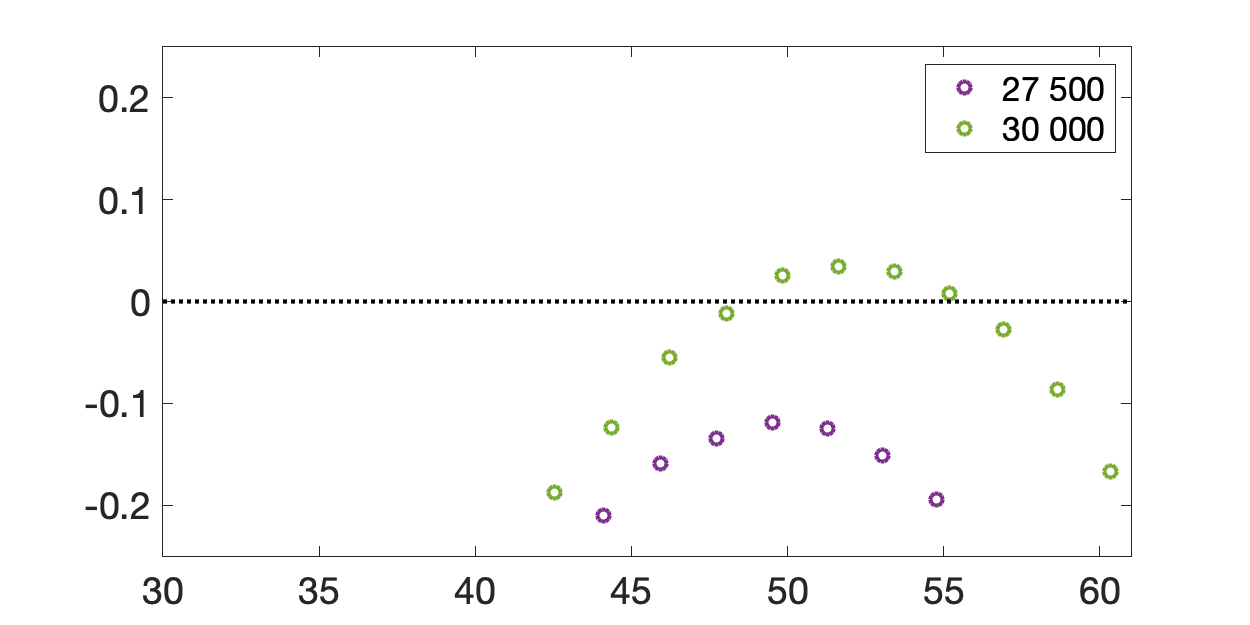}
&
\includegraphics[width=.48\linewidth]{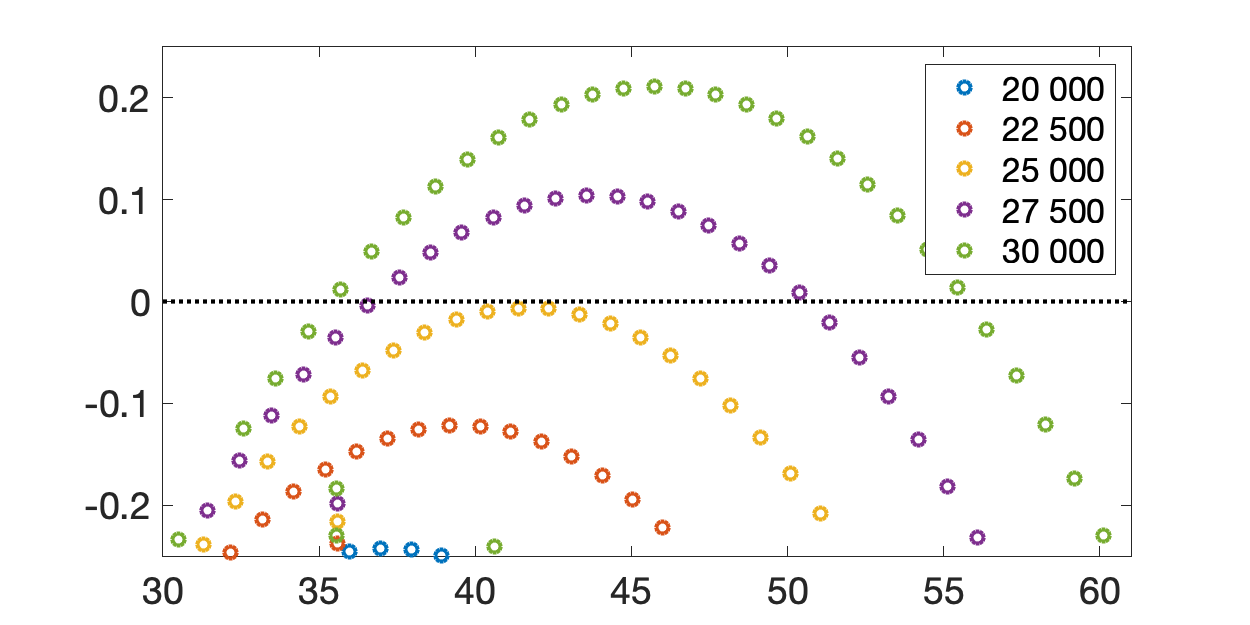}
\end{tabular}
\caption{Linear spectrum for the flow over a thin plate in a truncated domain, for several values of the Reynolds number.
$(a)$ : with domain length $L_x = 2.5$; $(b)$ with domain length $L_x = 5$.}
\label{fig:Plate_Spectrum}
\end{figure}

The case of a flat plate in a truncated domain, with the same parameters and same mesh as in the previous section, is now investigated 
using global stability analysis. Figure \ref{fig:Plate_Spectrum} shows the computed spectra (set of eigenvalues $\omega$ in the complex plane) for different values of $Re$ and $L_x$. As can be observed the spectra take the form of a very characterized arc-branch of eigenvalues.
For the shortest domain $L_x= 2.5$ (plot \ref{fig:Plate_Spectrum}$a$), all modes along the arc-branch are found to be stable for $Re = 27\,500$, while 
four unstable modes have emerged for $Re = 30 \, 000$. This feature is coherent with the time-stepping simulation discussed above, which indeed indicated destabilization between these two values of $Re$. For the longest domain $L_x= 5$ (plot \ref{fig:Plate_Spectrum}$b$), the spectra indicate that destabilisation occurs at a lower value of the Reynolds number, namely very close to $Re = 25 \,000$, the value a which the arc-branch emerges on the unstable half-plane. Note that both cases $Re = 27 \,500$, and $Re = 30 \,000$ are found to be strongly unstable, with a large number of eigenvalues (up to 21 for the highest value of $Re$) in the half-plane $\omega_i >0$.

The results presented here display all the typical features of the arc-branch spectra observed in other shear flows, for instance those 
the inhomogeneous jets considered by \cite{coenen2017global} and \cite{chakravarthy2018global}. In particular, the clustering of the eigenmodes along the arc-branch is a typical trend already pointed out in these studies. The spacing between the eigenvalues along the branch, noted $\Delta \omega_r$, is observed to be inversely proportional to the domain length, namely  $\Delta \omega_r \approx L_x^{-1}$, indicating that the domain length is controlling the discretization along the arc-branch. Note also that the existence of a large number of unstable eigenmodes can be related to the observation of modulation in nonlinear simulations, which can be interpreted as a competition between several modes.

\begin{figure}
\includegraphics[width=.99\linewidth]{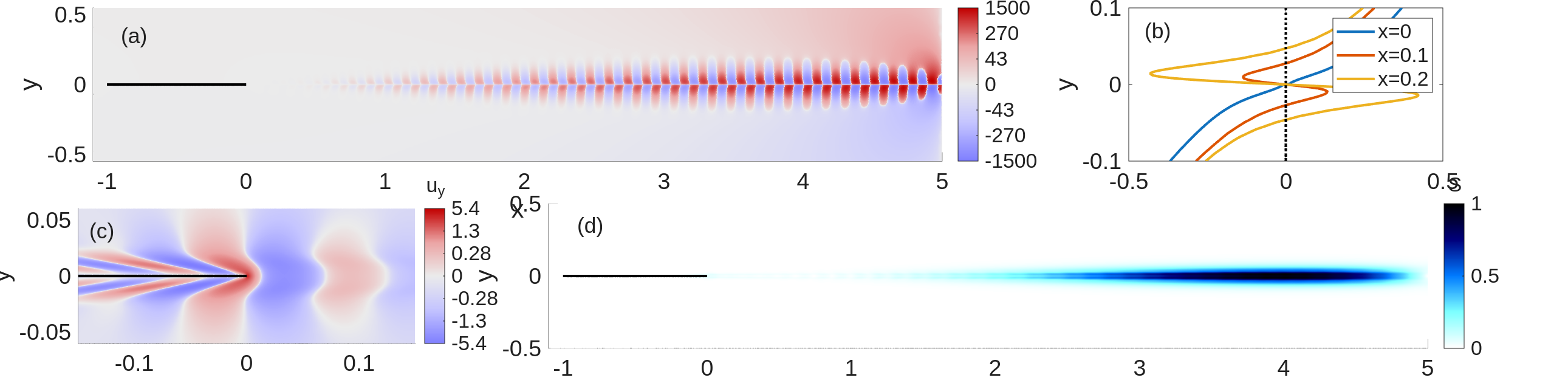}
\caption{Description of one of a near-neutral mode of a thin plate, computed for $Re = 25 \, 000$, with domain length $L_x = 5$. $(a)$: iso-contours of the pressure component $\tilde{p}(x,y)$; note the logarithmically stretched color map; $(b)$ detail of the pressure in the vicinity of the trailing edge. $(c)$ adjoint eigenmode (transverse velocity component), on a range centered around the trailing edge ; $(d)$ Structural sensitivity field $S(x,y)$.   }
\label{fig:Plate_Mode}
\end{figure}

To elucidate the paradoxal emergence of global instability in a flow which is expected to be stable owing to the absence of region of absolute instability, let us now have a look at the structure of the corresponding eigenmodes. Figure \ref{fig:Plate_Spectrum} displays the structure of one of the marginally unstable eigenmodes computed for $L_x = 5$ and $Re = 25\, 000$ 
Plot \ref{fig:Plate_Spectrum}$(a)$ displays the pressure component $\tilde{p}(x,y)$ of the global eigenmode. The figure allows to detect quite clearly two components in this pressure field. First, a wave-like component which is spatially amplified all along the domain from the trailing edge ($x=0)$ down to the numerical outlet ($x=L_x=5$). Secondly, one observes a second component which is not wave-like, but rather looks like a quadripolar field generated at the location where the wake encounters the numerical outlet. 

Note that the amplitude of the eigenmode has been normalized assuming by imposing $\tilde{u}_y(0.1,0) = 1$, implying that the amplitude in the vicinity of the trailing edge is of order one (as also indicated by plot \ref{fig:Plate_Spectrum}$(b)$ which details the structure of the pressure field in this region). In contrast, the values of the pressure perturbation at  the location of the outer boundary is more than 1000 times larger, indicating that the convective amplification along the wake is huge. This huge range of amplification lead us to use logarithmically stretched color map in plot \ref{fig:Plate_Spectrum}$(b)$ ; otherwise nothing but the few last wavelengths of the convective structure would have been visible in the figure.

Along with the structure of the eigenmode, it has also became a usual practice in global stability analysis to consider the adjoint eigenmode and the structural sensitivity \citep{Chomaz2005,giannetti2007structural} which both give complementary informations regarding the mechanism responsible for instability. The adjoint eigenmode can be physically interpreted as the initial perturbation which has the largest projection onto the eigenmode and hence leads to the largest transient nonnormal amplification. With no surprise, the adjoint eigenmode is strongly localized in the vicinity of the trailing edge, hence only a zoom of this region is displayed in figure \ref{fig:Plate_Spectrum}$(c)$.

The structural sensitivity $S(x,y)$ is a quantity initially introduced by \cite{giannetti2007structural}, mathematically defined as the product of the norms of the eigenmode and its adjoint : $S(x,y) = | \tilde{\bf u} | \cdot   | \tilde{\bf u}^\dag |$. Physically, this quantity is a marker of the region(s) where modifications of the linear operator have the largest effect on the eigenvalues ; also called the "wavemaker(s)" regions. 
Focussing again on the same near-neutral eigenmode as considered so far, the corresponding structural sensitivity is plotted on figure \ref{fig:Plate_Spectrum}$(d)$. The plot displays a characteristic streak-like structure extending all the way along the wake and reaching its maximum value closely before the outlet of the numerical domain.
This structure is very different from the one usually encountered in the case of blunt bodies, which is strongly localized within the recirculation region, in accordance with the argument based on absolute instability which also points towards this region as the origin of the instability mechanism. In contrast, the structure evidenced here suggests that the vicinity of the trailing edge play little role in the selection of frequency, while the vicinity of the outlet boundary is essential.

As a final remark, although a single specimen among the numerous nearly-neutral eigenmodes encountered in the spectra
 was displayed in figure \ref{fig:Plate_Spectrum}, all other eigenmodes actually look almost identical. The only difference between eigenmodes 
is the number of wavelengths of the convective structure from the trailing edge to the outlet. For instance, the eigenmode displayed in the figure, which correspond to $\omega = 42.35 -0.007 i$, has 36 wavelengths ; its closest neighbours along the arcbranch, corresponding to  $\omega = 41.37 -0.008 i$
$\omega = 43.33 -0.013 i$, have respectively 35 and 37 wavelengths.

\subsection{Hints from local stability theory ?}

As explained in \S 2.3, local stability is a simpler framework which is particularly relevant to describe the spatial amplification of disturbances in situations  where the base flow is nearly parallel; or more precisely when the characteristic length of variation of the base flow is larger than the wavelength of the disturbances. The inspection of the base flow (figure \ref{fig:PlateBaseFlow}) has shown that this hypothesis is well verified, so this approach is carried out here. 

\begin{figure}
\includegraphics[width=.49\linewidth]{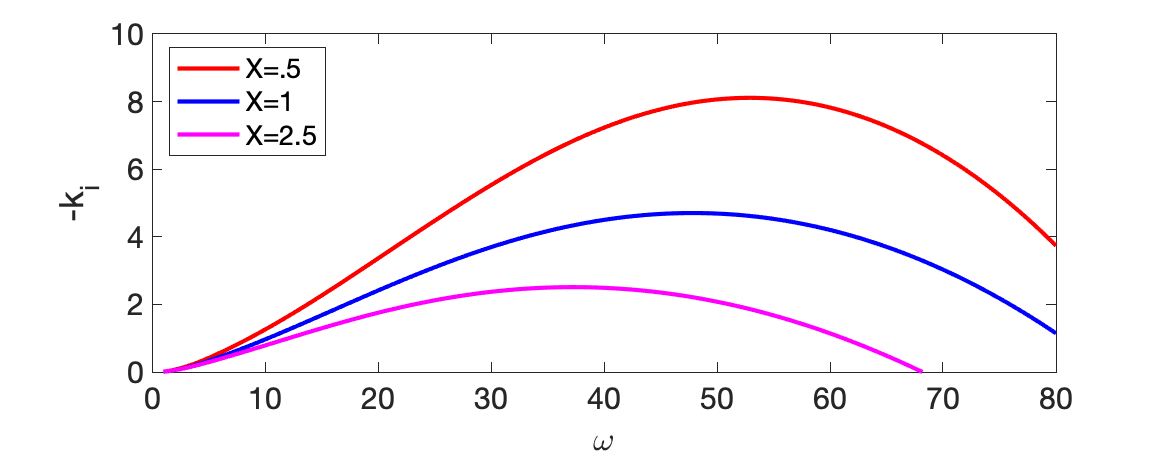}
\includegraphics[width=.49\linewidth]{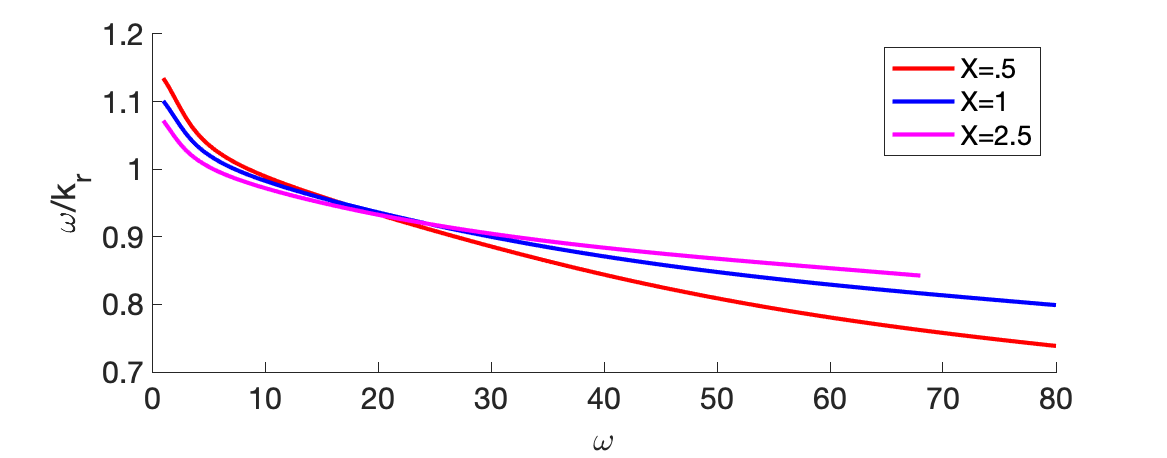}
\caption{Local (spatial) stability results for parallel wake profile extracted from the base flow for $Re = 27\, 500$ at several values of $x$. }
\label{fig:LocalStab}
\end{figure}

Figure \ref{fig:LocalStab} displays results from the local stability theory taking as a base flow the axial velocity profiles extracted from figure \ref{fig:PlateBaseFlow}$(b)$, at several locations of the downstream distance $x$, again with $Re = 27\,500$. Here a spatial stability framework was adopted, which consist of solving for complex values of $k$ considering real values of $\omega$. This framework is more relevant to describe the structure of nearly-neutral global modes. The left plot shows the spatial amplification rate $-k_i$ as function of the frequency. 
This quantity
reaches a maximum for values of $\omega$ in the range $[40-50]$. As $x$ is increased and one moves downwards in the wake, the amplification rate becomes smaller,  while the most amplified frequency shifts towards smaller values of $\omega$. The first trend is a consequence of the weaker intensity $R$ of the wake, while the latter is explained by he diffusion of the wake profile, as an increase of the thickness $\delta$ is correlated to a increase of the local wavelengh of the most amplified perturbation, and hence a decrease of the frequency. The right plot shows the phase velocity ($k_r/\omega$) of the  perturbation. In the range of $\omega$ corresponding to maximal amplifications, this group velocity is about $85 \%$ of the incoming velocity, a value comparable to that of the axial velocity along the axis ($y=0$) at the corresponding axial positions.

As already noted, in most of the wake, the base flow can be accurately fitted with a Gaussian law. The local stability of Gaussian wakes is a classical problem treated in reviews and textbooks (see, e.g. \cite{HuerreMonkewitz1990}) and the results presented here are fully in line with these classical results.
Note that for the axial locations considered in the figure, the instability is strongly convective and only a branch belonging to the $k^+$ family is displayed. Branches of the $k^-$ family are also present in the spectrum, but they are located far away in the spectrum, indicating that perturbations propagating upstream decay almost immediately. 
Hence such $k^-$ branches cannot be evoked to explain the feedback mechanism leading to self-sustained modes. An alternative explanation, featuring non-local perturbations which are not describable by the local approach, will be presented in section 4.

Finally,  the absence of any $k^-$ branch also confirms the already announced absence of any region of absolute instability, as the latter usually emerges from a well-known pinching process involving two branches of the $k^+$ and $k^-$  families. This conclusion was already drawn from the absence of a recirculation region. This point has to be precised here, as the existence of a recirculation region is not a strict criterion for absolute instability. Indeed, for gaussian wakes, absolue instabilities arises when $R>0.95$, meaning that the velocity along the axis $U(0)$ needs not to be negative but only smaller than $5\%$ of the outer velocity. A region where $U(0)<0.05$ is present here in the very close vicinity of the trailing edge (for $x<0.05$). However in such regions the wake is not of gaussian shape, and local stability conducted with the local velocity law instead of the gaussian fit do not yield any absolute instability, in agreement with the conclusions of \cite{taylor1999note}.

Before closing this section, let us discuss what can be learned from the local stability framework regarding the structure of the adjoint eigenmode and the structural sensitivity. When searching solutions with form $[\hat{u}_x^\dag, \hat{u}_y^\dag, \hat{p} ^\dag] e^{i k^\dag x - i \omega^\dag t}$ of the adjoint   of the general linearized problem \ref{EigenvalueProblem}, in the local framework one can reduce the problem to the adjoint Orr-Sommerfeld equation:
 \be{OSA}
{\cal L}^\dag_{OS} ( \hat{u}_y^\dag ) \equiv (k^\dag U- \omega^\dag ) \left( \partial_y^2 - {k^\dag}^2\right)  \hat{u}^\dag_y  + 2 k U' \partial_y \hat{u}^\dag_y + i Re^{-1} \left( \partial_y^2 - {k^\dag}^2\right)^2  \hat{u}^\dag_y 
 = 0
 \ee
The relationships between the direct and adjoint Orr-Sommerfeld problems were deeply analyzed in \cite{salwen1981continuous} and \cite{hill1995adjoint}. An important consequence of these works is to show that, in the spatial framework, if $(\omega,k_r+i k_i)$ is a solution of the direct problem, then $(\omega^\dag, k^\dag) = (\omega, k_r - i k_i)$ is a solution of the adjoint problem. This signifies that the spatial damping rate $k_i^\dag$ of the adjoint mode is opposite to the amplification rate $(-k_i)$ of the direct eigenmode. A consequence is that in the local stability framework, the structural sensitivity, defined as the product of direct and adjoint modes, is constant and independent upon the spatial loction $x$. This provides a clear explanation to the long streak of structural sensitivity within the wake, as observed in figure \ref{fig:Plate_Mode}$(d)$.

\section{Nonlocal feedback model in the linear regime}

As a conclusion of the previous section, the local stability framework, assuming a parallel flow $U(y)$ occupying an infinite domain $x\in [-\infty, \infty]$, correctly describes the downstream amplification of perturbation but is unable to predict any feedback mechanism necessary for the existence of self-sustained oscillations. The purpose of this section is to derive a model predicting such a feedback, by reconsidering the problem of a parallel flow occupying a domain bounded in in the downstream direction at the location of the outlet of the numerical domain $(x=L_x)$.

\subsection{Nonlocal modeling of the pressure}

One starts by investigating the effect of domain truncation by considering eigenmode solutions of the linearized problem, again under the form $[\tilde{\bf u}, \tilde{p}] e^{-i \omega t}$, within a  semi-infinite domain $(x,y) \in [-\infty, L_x] \times [-\infty,\infty]$ where $L_x$ is the location of the outer boundary. The velocity field of the eigenmode is still assumed as a {\em wave component} given by the {\em local eigenmode} ansatz :
\be{ansatzNLocveloc}
\tilde{\bf u}(x,y) e^{-i \omega t}  = a_0 \hat{\bf u}(y)  e^{ik x} e^{ - i \omega t}.
\ee 
Where both $\omega$ and $k$ are a priori complex. Without loss of generality, one can normalize the eigenmode is such a way that the transverse velocity component along the axis is unity, i.e. $\hat{u}_y(y=0) = 1$. With this normalization, the amplitude $a_0$ corresponds to the value of the transverse velocity at location of the trailing edge ($x=0,y=0$).

Consider now the pressure. The Poisson equation \ref{Poisson}  leads to :
\be{LaplPLin}
\Delta \tilde{p} 
= -4 i \ddp{U_{0}}{y} \hat{u}_y
  a_0 k e^{ikx}
\quad \mbox{ for } x<L_x;
\qquad p(x=L_x, y) = 0. 
\ee
This is a Poisson equation for the pressure $\tilde{p}$ in the half-plane $x<L_x$, along with a Dirichlet condition at the right boundary. One may extend this equation on the whole plane by assuming antisymmetry of both $p$ and the source term with respect to the line $x=L_x$. The resulting problem is:
\be{LaplPLin2}
\Delta \tilde{p} =  \tilde{S}_p(x,y) = \left\{
\begin{array}{cc} a_0 k \hat{S}_p(y) e^{ikx} &\quad \mbox{ for } x<L_x\\
- a_0 k \hat{S}_p(y) e^{ik(2L_x-x)}  &\quad \mbox{ for } x>L_x
\end{array}
\right. \quad \mbox{ where } {\hat{S}_p(y)} = -4 i \ddp{U_{0}}{y} \hat{u}_y
\ee

This Poisson equation can be solved using Green's method as follows

\begin{eqnarray}
\tilde{p}(x,y)  &=&   \int\limits_{-\infty}^{+\infty}   
\int\limits_{-\infty}^{\infty}  
\tilde{S}_p(x',y')
\frac{\log|{\bf r'-r}|}{2 \pi}L_x dx'dy' 
\label{LaplPLin}
\\
&=&  a_0 k  \int\limits_{-\infty}^{+\infty}  {\hat{S}_p(y')}  \left[ 
\int\limits_{-\infty}^{L_x}  
\frac{\log|{\bf r'-r}|}{2 \pi}
e^{ikx'}  dx' 
- \int\limits_{L_x}^{+\infty} 
\frac{\log|{\bf r'-r}|}{2 \pi}
e^{ik(2L_x-x')} dx' \right] dy' 
\nonumber
\end{eqnarray}
Where one recognizes the Green function associated to the Laplacian operator on an infinite two-dimensional domain :
$G_0 = \frac{\log|{\bf r'-r}|}{2 \pi} = \frac{\log\left[ \sqrt{(x-x')^2 + (y-y')^2 }\right]}{2 \pi} $.

This expression can can be re-writen by adding an substracting the integral of the first term over the fictitious domain $x\in[L_x,\infty]$, leading to an expression of the form

\be{LaplPLin2}
\tilde{p} =  \tilde{p}^{(loc)} + \tilde{p}^{(Nloc)} 
\ee
with 
\be{ploc}
 \tilde{p}^{(loc)} =  a_0 k \int_{-\infty}^{+\infty}  \int_{-\infty}^{+\infty}  {\hat{S}_p(y')} \frac{\log|{\bf r'-r}|}{2 \pi}
  e^{ikx'} dy' dx'
\ee
\be{pNloc}
 \tilde{p}^{(Nloc)} =   -a_0 k \int_{L_x}^{+ \infty}  \int_{-\infty}^{+\infty}  {\hat{S}_p(y')} 
 \frac{\log|{\bf r'-r}|}{2 \pi}
 \left[ e^{ik(2L_x-x')} + e^{ikx'}\right] dy' dx'
\ee

Under this form, the pressure is split into two components which have a clear meaning.
The first component is the {\em local} part of the pressure. Indeed one can verify that this part has modal dependance upon $x$ and can be written as $ \tilde{p}^{(loc)} = a_0 \hat{p}(y) e^{i k x}$, where $\hat{p}$ is the pressure component of the eigenmode with characteristics $(k,\omega)$ predicted by the local approach of the previous section. 

The second  component is the {\em non-local} part of the pressure and has a priori no modal dependance over $x$.
To close the feedback loop in the next subsection  an approximation of this term in the vicinity of the trailing edge, i.e. $x\approx 0$, $y \approx 0$, is needed. In this case two approximations can be invoked to simplify this term. First, one can remark that the source term $\hat{S}_p(y')$ in the pressure equation is an antisymmetric function of $y'$ with a compact support localized close to $y'=0$. This can be approximated by a doublet singularity : 
$\hat{S}_p(y') \approx K_p \delta'(y')$ where $\delta'$ is the derivative of the Dirac function and $K_p$ some scaling factor. Under this hypothesis the expression becomes a  1-D integral along the single variable $x'$, with the form
\be{pNloc1}
 \tilde{p}^{(Nloc)} \approx - a_0  \int_{L_x}^{+ \infty}  
  \frac{K_p k y}{2 \pi  \left[ (x'-x)^2 + y^2\right]} 
 \left[ e^{ikx'} + e^{-ik(2L_x-x')}\right] dx' 
\ee
Where one recognizes
$\displaystyle G^1= \left.\frac{\partial G^0}{\partial y'}\right|_{y'=0} 
=  \frac{y}{2 \pi  \left[ (x'-x)^2 + y^2\right]} 
$ as the Green function associated to a doublet singularity. 
Redefining the variable of integration as $x' = x''+L_x$ , one is now led to:
\be{pNloc1}
 \tilde{p}^{(Nloc)} \approx a_X \int_{0}^{+ \infty}   \frac{-K_p k y}{\pi  \left[ (x''+L_x-x)^2 + y^2\right] } \left[ e^{ikx''} + e^{-ikx''}\right] dx'' 
\ee
where $a_X = a_0 e^{i k L_x}$ is the amplitude of the convectively unstable local eigenmode at the location $x=L_x$ of the outlet. 

For the next step of the model, an approximation of this nonlocal pressure away from the outlet but close to the axis, i.e. for $y \ll 1$ and $(L_x-x) \gg 1$, is needed. 
In this limit, the integral above can be expressed  in terms of the exponential integral function $E_i(x)$, and one can use the leading order approximation of this function to finally obtain

\be{pNlocAsympt}
\tilde{p}^{(Nloc)}(0,y) 
\approx a_X \frac{ -4 i K_p}{\pi k } \frac{y}{|L_x-x|^3}  
\ee

This important result shows that the nonlocal part of the pressure decays {\em algebraically} when moving away 
from the location of the outlet of the domain. More precisely, the algebraic decay obtained here is characteristic 
of a quadripolar pressure source. Such a structure was already identified in figure \ref{fig:Plate_Mode}$(a)$ as a constitutive part of the pressure field of the eigenmode. 


\subsection{Receptivity model and discretization condition}

To close the problem, one needs a "receptivity model" linking the initial amplitude of the unstable wave to a forcing by the nonlocal pressure. Indeed, inspection of the pressure component of the 
eigenmode (see plot \ref{fig:Plate_Mode}$b$) shows that the nonlocal pressure is, despite of its algebraic decay, still present in the vicinity of the trailing wake.
Physically, one can assume that the oscillations of the wake will be triggered by a gradient of this pressure in the transverse direction, 
namely $\tilde{u_y}(0,0) = K_r \left. \ddp{\tilde{p}}{y}\right|_{y=0}$ where $K_r$ is a receptivity coefficient.
This equation and the one accounting to the downstream amplification if the wave component lead to two conditions linking the amplitude $a_0$ at the trailing edge and the amplitude $a_X$ at the outlet of the domain, namely
$$
a_X = a_0 e^{i k L_x}, \qquad a_0 = \frac{K}{ k L_x^3} a_X \qquad \mbox{ with }  \quad K = \frac{4 i K_p K_s}{\pi}
$$
This eventually leads to the discretization condition as follows :
\be{discretizationLin}
e^{i k L_x} \frac{K}{k L_x^3} = 1
\ee 
This discretization condition admits a discrete set of solutions for the axial wavenumber $k$ of the wave component,
as function of the size of the domain $L_x$ in the downstream distance, and of a single factor $K$ which regroups the fine details of the models including the receptivity condition. In the following, the value of this parameter is set to $K=5 \cdot 10^{-4}$, a value which was found to correctly reproduce the global stability results.

\subsection{Resolution}

\begin{figure}
$$
\includegraphics[width=.49\linewidth]{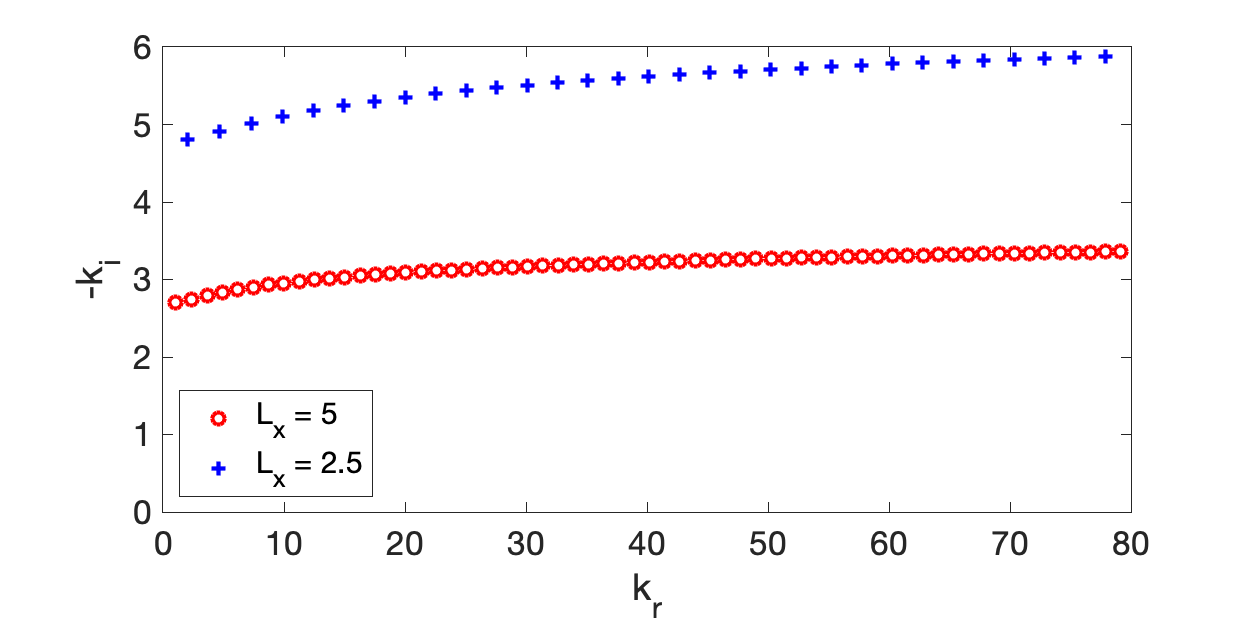}
\includegraphics[width=.49\linewidth]{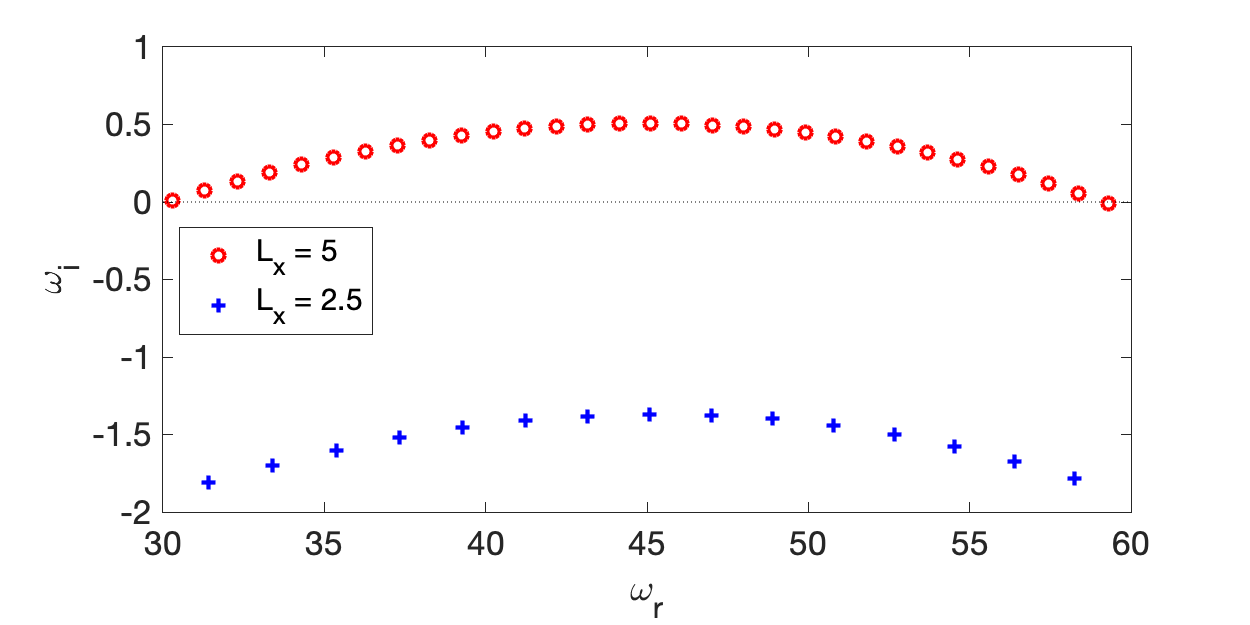}
$$
\caption{Predictions of the non-local feedback model. $(a)$ : Solutions of the discretization condition \ref{discretizationLin} in the complex $k$-plane, for $L_x= 2.5$ and $L_x = 5$.  $(b)$ : corresponding values of $\omega$  deduced by applying the local stability analysis, using the velocity profile at location $x=1$ as a parallel base flow.
} 
\label{fig:NlocLin}
\end{figure}

Figure \ref{fig:NlocLin}$(a)$ shows the solutions in the complex $k$-plane for two domain sizes corresponding to $L_x= 2.5$ and $L_x=5$. In both cases one obtains a discrete set of values, all characterized by a negative imaginary part. Note also that the real part of $k$ are regularly spaced, and that their spacing is inversely proportional to $L_x$. This is consistent with the fact, already identified in the previous section, that the wave component of the eigenmode has an integer number of wavelength along the domain.

Once the discretization condition solved for the wavenumber $k$, one can deduce the corresponding frequency $\omega$ or the global eigenmode by solving the local stability problem in a spatio-temporal framework (solving for a complex $\omega$ as function of a complex $k$). This has been done by selecting the velocity profile extracted from the steady flow solution at axial distance $x=1$ and $Re = 27 \, 500$ (taken from figure \ref{fig:PlateBaseFlow}$b$).The choice $x=1$ is done with the idea of considering an "average" of the instability properties of the wake flow over the domain. Results obtained in this way are plotted in figure \ref{fig:NlocLin}$(b)$. As can be seen, the model leads to a well characterized arc branches which reproduce very well the ones observed by global stability analysis. Especially, with the chosen value of $K$, the problem gets instable when the domain size is increased from 2.5 to 5, just as in global results, and the increased clustering of the modes along the arc branch is also accurately reproduced.




\section{The NACA0012 profile with zero incidence} 

\subsection{Time-stepping simulations on a truncated domain}

\begin{figure}\centering
\begin{tabular}{c}
\begin{tabular}{cc}
\includegraphics[width=.49\linewidth]{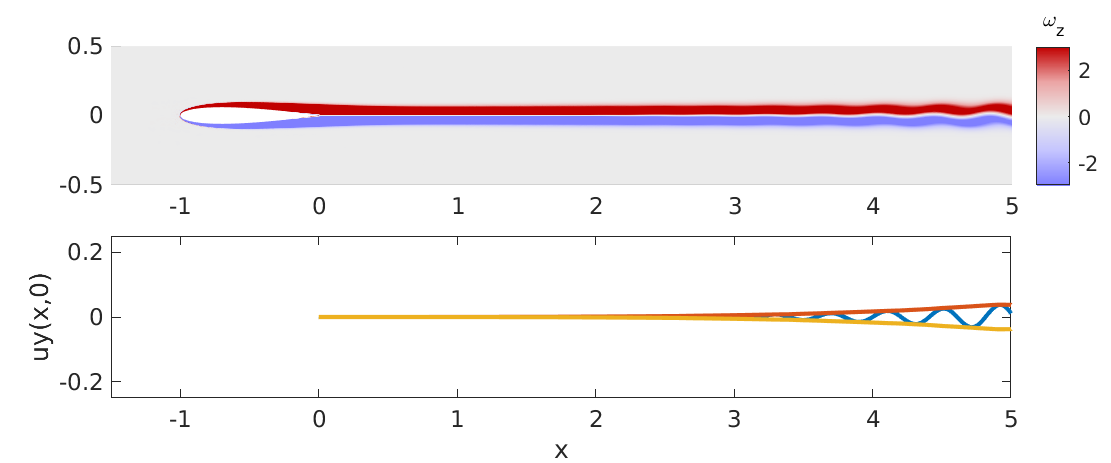} 
&
\includegraphics[width=.49\linewidth]{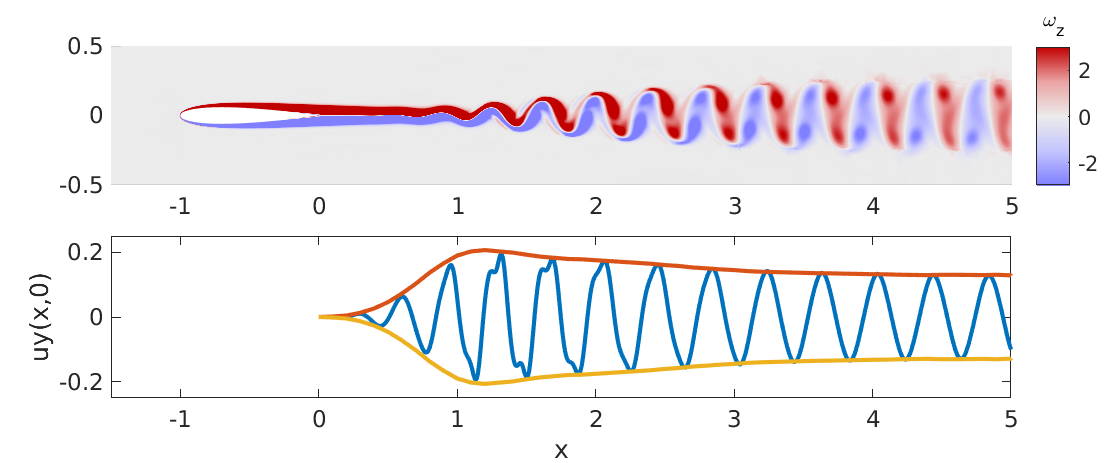} 
\\ 
$(a)$ & $(b)$
\end{tabular}
\\
\includegraphics[width=.95\linewidth]{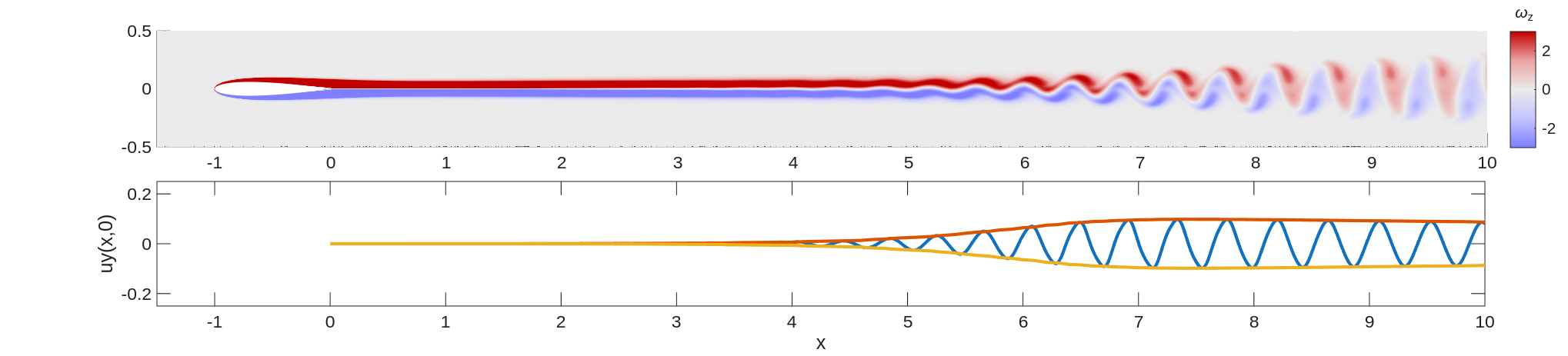}
\\
$(c)$
\\
\includegraphics[width=.95\linewidth]{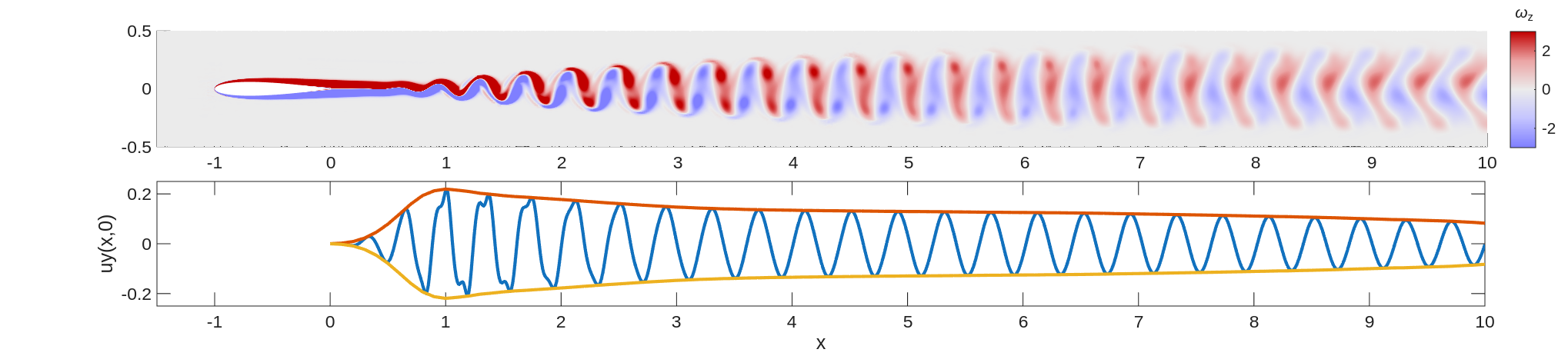}
\\
$(d)$
\end{tabular}
\caption{DNS snapshots (color plot of vorticity) and corresponding profiles of velocity along the axis. 
$(a)$ : $L_x = 5,Re = 7500$ ; $(d)$ : $L_x = 5,Re = 9000$ ; $(c)$ : $L_x = 10,Re = 7000$ ; $(d)$ : $L_x = 10,Re = 9000$.
}
\label{fig:Naca12_DNSSnapshots_L5}
\end{figure}

%
%

Figure \ref{fig:Naca12_DNSSnapshots_L5} shows DNS snapshots (vorticity) along with profile of the velocity components (the thick line corresponds to the instantaneous value at the same time as the snapshot, while the colored lines indicate the envelope of the oscillating values over a full cycle) for selected values of the domain size $L_x$ and the Reynolds number $Re$.

Consider, first, a rather short truncated domain, namely $L_x = 5$. In this case, unsteadiness is observed to above a threshold estimated as $Re = 7,400$. The flow for $Re = 7500$, just above this threshold, is displayed in plot \ref{fig:Naca12_DNSSnapshots_L5}$(a)$. Just like the thin plate, unsteadiness is observed to first arise in the vicinity of the domain outlet, raising the question of a possible spurious feedback from the boundary condition.
For $Re \ge 9,000$, (plot \ref{fig:Naca12_DNSSnapshots_L5}$b)$ a robust vortex shedding phenomenon is observed, leading to a vortex street which saturates at a close distance from the trailing wake. 

Repeating the simulation on a longer domain, namely $L_x = 10$ , leads to similar observations. The main difference is that the onset of unsteadiness is observed to occur significantly earlier, with a threshold estimated as $Re=6900$. Slightly above this threshold, for $Re=7000$ (plot \ref{fig:Naca12_DNSSnapshots_L5}$c)$ the maximal amplitudes are again observed in the vicinity of the outlet, raising the question of a possible spurious effect. On the other hand, when increasing the Reynolds, the result become rather independent upon the size of the domain.
As an illustration, the flow for $Re = 9000$ (plot \ref{fig:Naca12_DNSSnapshots_L5}$d)$ is almost identical to the one obtained with the shorter domain.

Noting that the critical Reynolds number significantly varies with domain size, from $Re \approx 7400$ for $L_x= 5$ to $Re \approx 6900$ for $L_x = 10$, one is naturally tempted to repeat simulations on even longer domains, hoping to find a value of $L_x$ large enough to get rid from any possible pollution from boundary conditions. When doing so, the simulations get very time-consuming, for two reasons. First, and quite obviously, because longer domains leads to increase of mesh size. Secondly, the required duration of simulations in physical time also need to be increased when the domain is enlarged. Especially, when decreasing $Re$ towards an hypothesized threshold value, one observes the occurence of slow modulations, quite similar to those observed for the thin plate and displayed in figure 
\ref{fig:DNS_Plate}. It is thus very difficult to know if such modulations are  a slowly decaying transient towards a final periodic state of a chaotic state. The purpose of the next sections will be precisely to demonstrate that enlarging the domain is not the right route to establish results relevant to the case of a wing "in an infinite domain".

\begin{figure}
$$
\includegraphics[width=.49\linewidth]{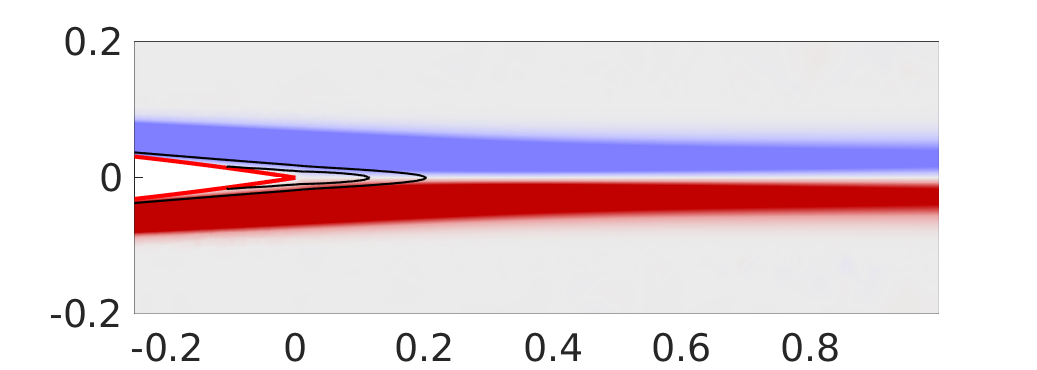}
\includegraphics[width=.49\linewidth]{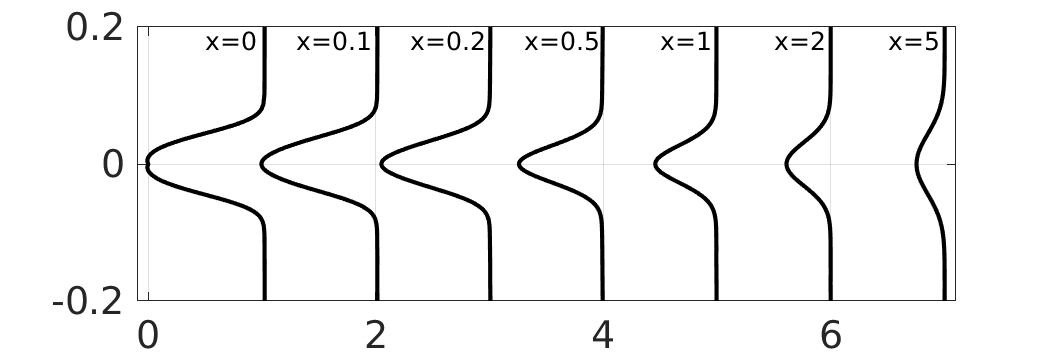}
$$
$$
\includegraphics[width=.49\linewidth]{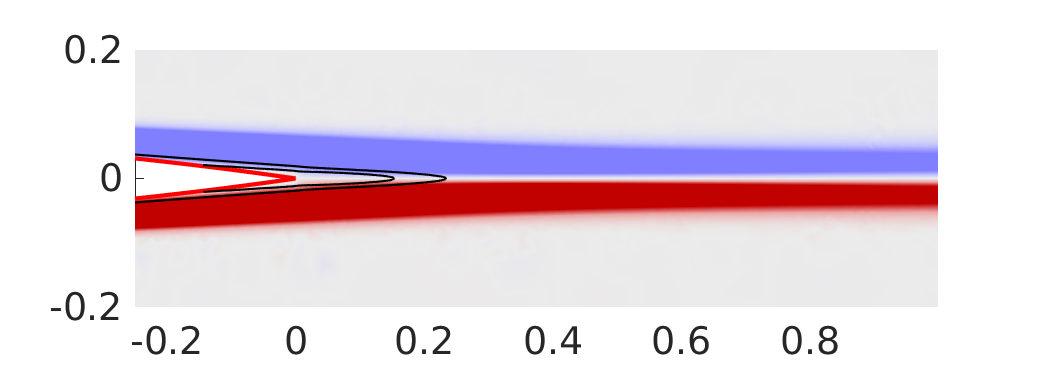}
\includegraphics[width=.49\linewidth]{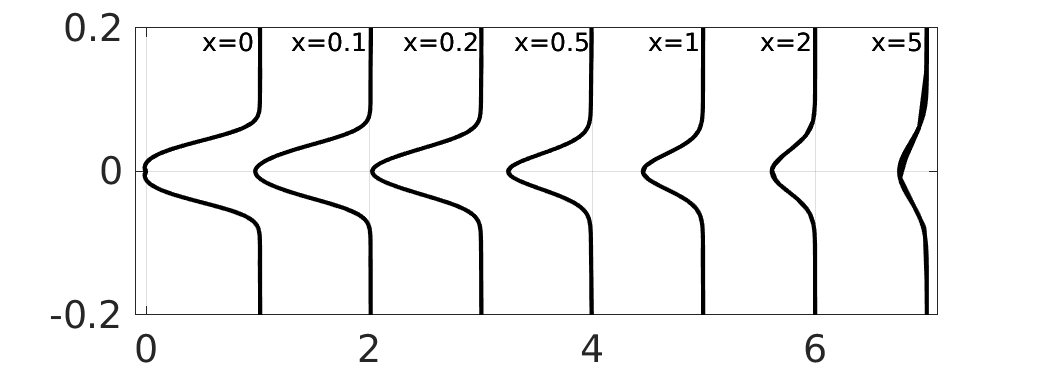}
$$
\caption{Structure of the base flow in the vicinity of the trailing edge (left), and profiles of axial velocity (right) at several axial locations $x$ (with an offset for better readability). Upper row : $Re = 6800$ ; lower row : $Re = 8500$.
In the left plots the two black lines are the isocontours $u_x = 0$ (recirculation region) and $u_x = 0.05$ (region of absolute instability).
} 
\label{fig:NACABase}
\end{figure}

Before closing this section, and just like for the plate, let us investigate the structure 
of the steady solution (base-flow) in order to identify possible "wavemaker" regions responsible for unsteadiness.
Figure \ref{fig:NACABase} displays the structure of the base flow for $Re = 6900$ (up) and $Re = 8500$ (down).
In both cases, a recirculation region is present (indicated by the black contour in the left plots), with comparable extension in both cases. Hence, local feedback through to the classical scenario linked to absolute instability can clearly happen here, but time-stepping results indicate that spurious non-local feedback from the boundary conditions may also be present.


\subsection{Global linear stability on a truncated domain}

\begin{figure}
\begin{tabular}{cc}
\includegraphics[width=.48\linewidth]{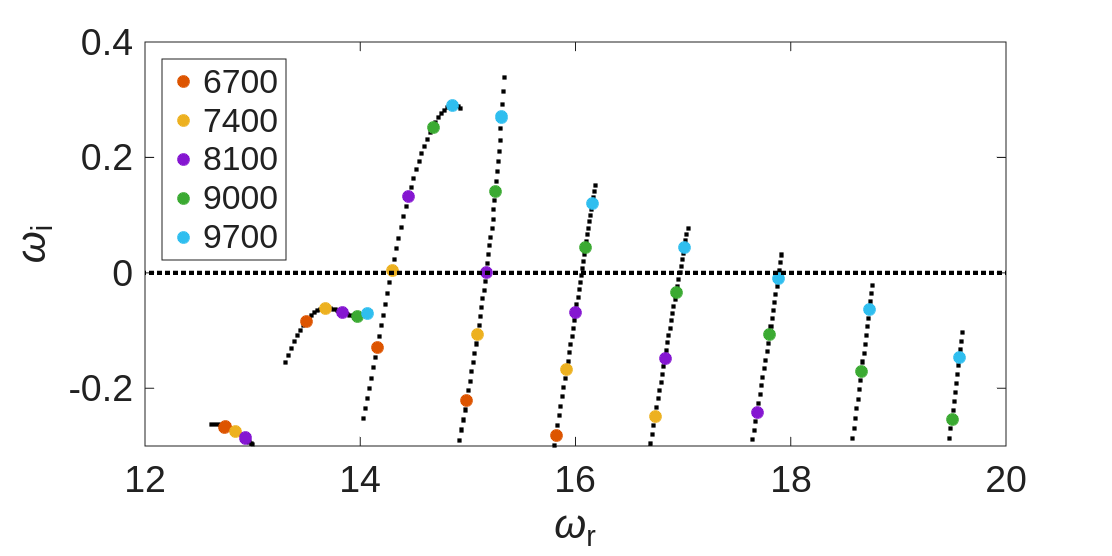}
&
\includegraphics[width=.48\linewidth]{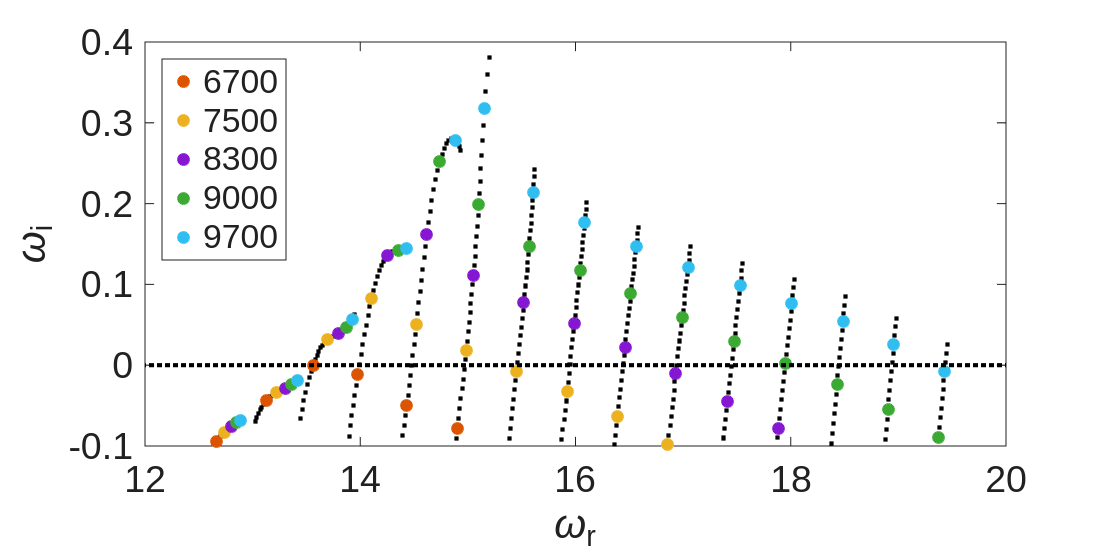}
\\
\includegraphics[width=.48\linewidth]{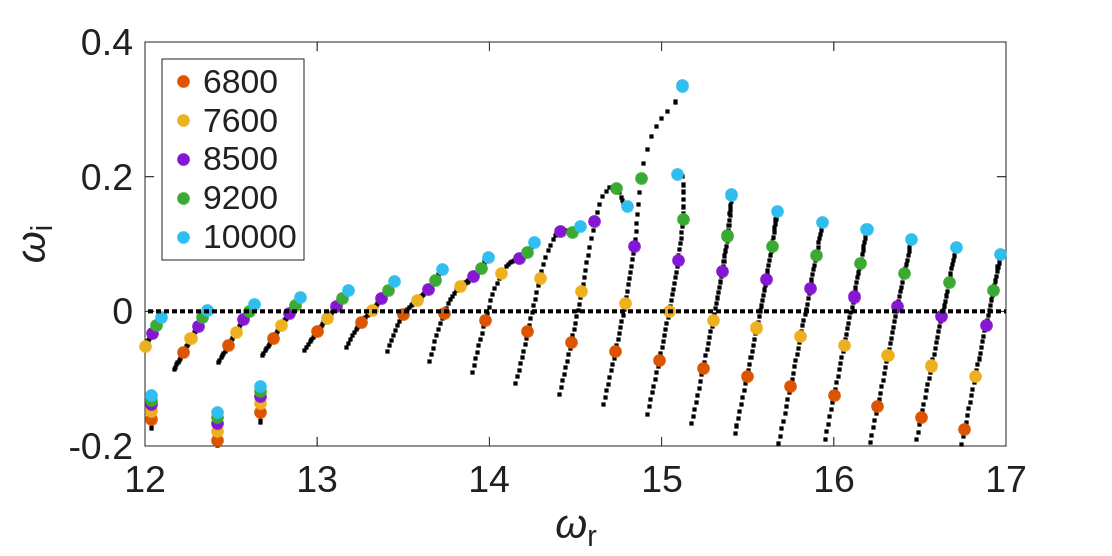}
&
\includegraphics[width=.48\linewidth]{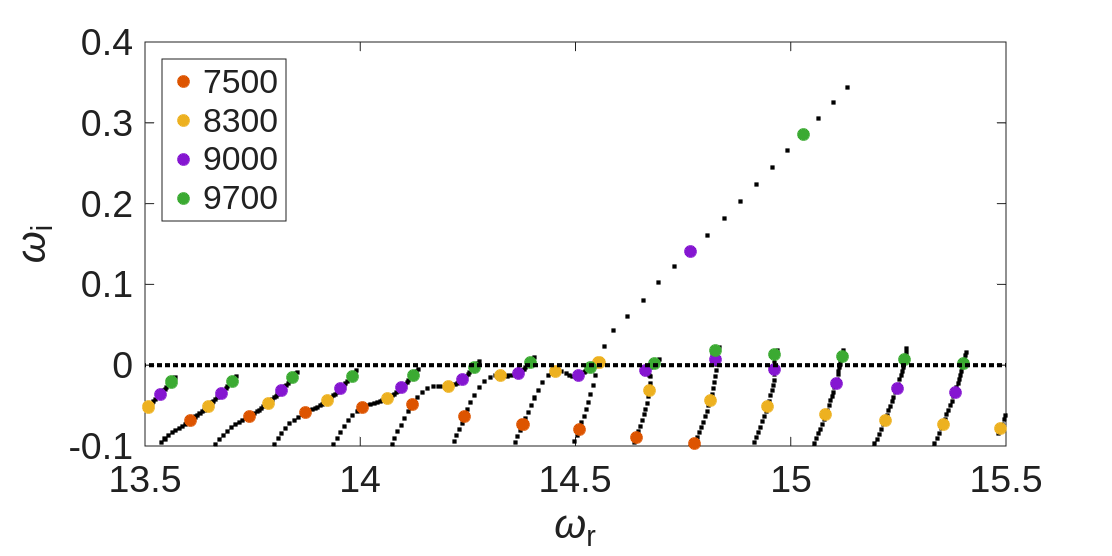}
\end{tabular}
\caption{Linear spectra for the flow over a NACA0012 on a truncated domain,  with respectively $L_x= 5$ $(a)$, $L_x= 10$ $(b)$, $L_x= 20$, $(c)$, and $L_x= 40$ $(d)$. The dots are spectra computed in the range $Re  \in [6\,000- 10\, 000]$, and the colored points highlight selected values of the $Re$.}
\label{fig:Naca12_Spectra}
\end{figure}

Let us now turn towards global stability analysis, which is a much more efficient tool compared to time-stepping when it comes to map instability thresholds. In this section, results are presented using a truncated domain, using the same meshes as for the time-stepping simulation previously displayed.
Figure \ref{fig:Naca12_Spectra} shows the spectra for four different values of the domain size. In the figures, the black dots are eigenvalues computed over the range of Reynolds number $Re \in [6\,000 - 10 \,000]$, and the colored symbols highlight a few specified values of $Re$. 

For $L_x= 5$ (plot \ref{fig:Naca12_Spectra}$a$), one observes a dozen of well-defined eigenvalues, with an organisation which is reminiscent of an arc-branch, although not as regularly organized than for the thin plate case.
The first emergence of an unstable mode is seen to occur for $Re \approx 7400$,  in good accordance with the first observation of unsteadiness in time-stepping simulation with the same domain size. One can also note that as the Reynolds number is increased, several secondary unstable eigenvalues emerge. The second occurs at 
$Re \approx 8100$ and ends up being more amplified than the first one. There are eventually up to 5 unstable modes for $Re = 10\,0000$.

\begin{figure}
\centering
\includegraphics[width = .99 \linewidth]{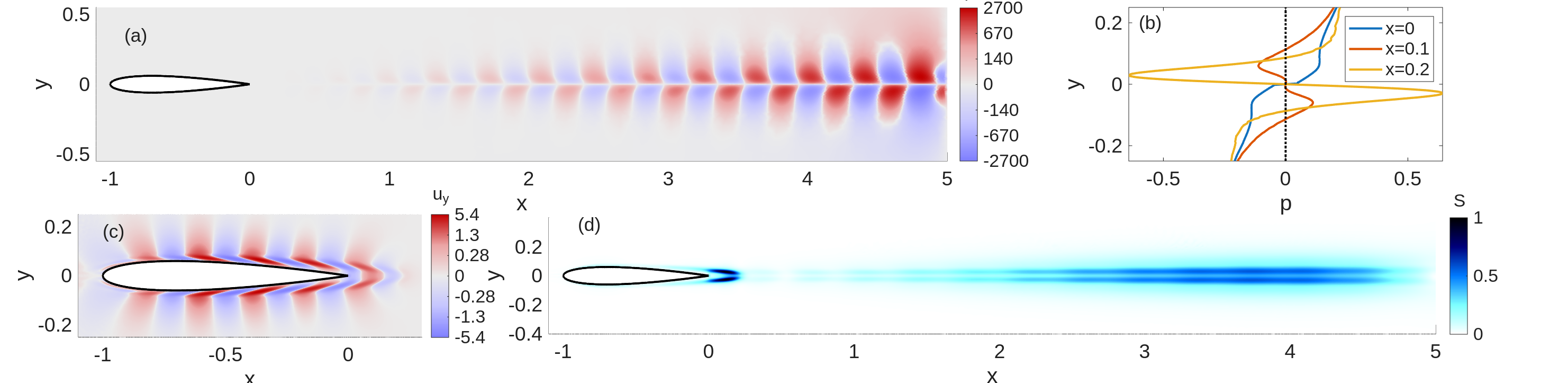}
\caption{Near-neutral eigenmode,  for $Re = 7400$, computed on a truncated domain with $L_x = 5$ along with corresponding adjoint and structural sensitivity. Same conventions as in figure \ref{fig:Plate_Mode}.
}
\label{fig:Naca12_Re7400_Eigenmode}
\end{figure}

Before discussing higher values of the domain size, let us depict the structure of the emerging eigenmode computed for $Re = 7400$. Figure \ref{fig:Naca12_Re7400_Eigenmode} displays this mode with the same conventions as used for the plate in figure \ref{fig:Plate_Mode}. Inspection of the pressure component (plot  \ref{fig:Naca12_Re7400_Eigenmode}$a$) show the presence of both a wave component and a nonlocal component, and the latter remains significant in the vicinity of the trailing edge (plot \ref{fig:Naca12_Re7400_Eigenmode}$b$).
Note also, as emphasized by the logarithmically streched color scale, that the wave component is amplified by a factor of more than 1000 between the vicinity of the trailing edge and the domain outlet.
These observation support the hypothesis that the nonlocal feedback mechanism evidenced in the previous sections is also present here. However, the structural sensitivity (plot  \ref{fig:Naca12_Re7400_Eigenmode}$d$) allows to identify two distinct active region. The first is the streak with almost constant values all along the axis, already observed for the plate, and again a hint for the presence of nonlocal feedback. The second region is localized along the edges of the recirculation region in the near wake. This latter structure is very similar to the one observed for blunt bodies driven by the absolute instability scenario. These results suggest that both local feedback through absolute instability and nonlocal feedback due to the boundary condition might be present here, and the objective in the next section will be to separate these two effects.   

Let us now come back to figure \ref{fig:Naca12_Spectra} and consider spectra computed using longer domains.
For $L_x= 10$ (plot \ref{fig:Naca12_Spectra}$b$), an arc-branch like structure is again observed.   The first emergence of an unstable mode occurs  for $Re > 6700$, with an oscillation rate close to $\omega_r \approx 13.5 i$. As the Reynolds is raised above this value, an increasing number of unstable eigenvalues are encountered. Note that the first branch to emerge is not the most amplified, as it is quickly overpassed by the few ones emerging subsequently ; for $Re = 10\, 000$ the most amplified is the fourth one to emerge, with an oscillation rate $\omega_r \approx 14.8$.

 Enlarging the domain length up to $L_x= 20$ (plot \ref{fig:Naca12_Spectra}$c$), shows a globally similar picture, with an increased clustering of eigenvalues along an arc-like structure. This time the instability first arises at a threshold $Re \approx 6800$, with the simultaneous onset of two unstable modes with oscillation rates $\omega_r \approx 13.50$ and $13.73$, respectively. Similarly as for $L_x=10$, as the Reynolds number is increased, the modes emerging subsequently quickly become more amplified than the first ones to emerge, and for $Re = 10\, 000$,  the most amplified one is again in the range $\omega_r \approx 14.8$.

At this stage, having observed that the threshold Reynolds number decreases when enlarging the domain but does not vary much between $L_x=10$ and $L_x= 20$, one may deduce that the latter case approaches the theoretical limit of the "infinite domain" , and that the critical Reynolds number is thus close to 6800. However, enlarging the domain size up to $L_x= 40$ (plot \ref{fig:Naca12_Spectra}$d$) leads to a completely different conclusion. In this case, the results reveal a single leading mode emerging for $Re \approx 8400$ and clearly dominating the spectrum above this value. Aside with this dominant mode, an arc-branch is also present. As expected this arc-branch is even more clustered than for smaller $Lx$ and a fraction of these modes emerge in the upper half-plane, but their amplification rate always remain much smaller than that of the dominating mode.

\begin{figure}
\includegraphics[width=.98\linewidth]{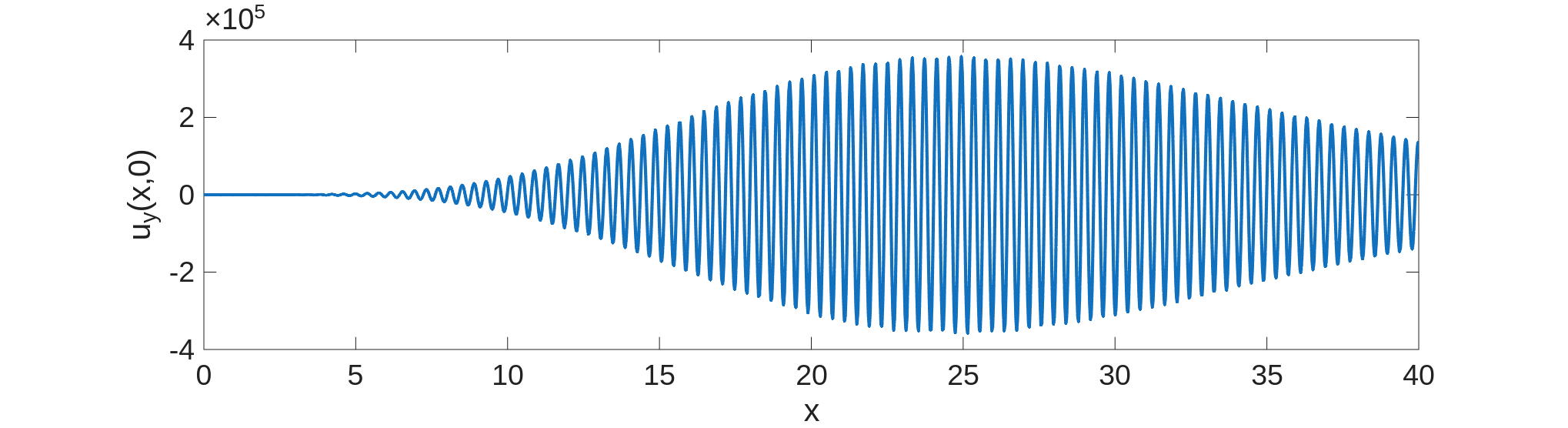}
\caption{Near-neutral eigenmode for $Re = 8300$ and $L_x = 40$.}
\label{fig:Naca12_Eigenmode8400}
\end{figure}

To explain why the global stability predictions keep on depending upon the domain size, let us have a look at the structure of the leading eigenmode with $L_x= 40$, $Re= 8400$, displayed in figure \ref{fig:Naca12_Eigenmode8400}. The structure is strongly elongated in the axial direction, thus it is not possible to represent the eigenmode with the same conventions as in the previous figures. Instead, the transverse velocity $\tilde{u}_y$ along the axis ($y=0$) is plotted as a function of $x$. 
The figure shows that the eigenmode reaches its maximum values for $x \approx 25$, explaining why computations done on shorter domains were necessarily strongly affected by transition. An other striking result of the figure is the incredibly huge level reached as a result of convective amplification. Recalling that, as for the previously displayed mode, the normalization is done by setting $\tilde{u}_y(0.1,0)=1$, meaning that amplitudes are of order one in the region of the trailing edge, an amplification of a factor larger than $10^5$ is observed.

Note also that although the amplitude of the mode decays between the peak at $x\approx 25$ and the outlet at $x = 40$ by about a factor 3, the amplitude at this location is still huge compared to the levels in the trailing edge region, suggesting that spurious feedback effects are still present here. At this stage one may be again tempted to repeat computations on longer and longer domains, but the next section will introduce alternative and much less costly ideas to tackle the idealized problem of a wake developing in an "infinite domain".


\subsection{Global stability analysis using downstream filtering}

\subsubsection{Principle}

Instead of extending the domain in hope of getting rid of spurious effect arising from boundary condition, which may require huge dimensions in the present case, a more rational idea is to design filtering methods suppressing the feedback. Three classes of methods are available for this, and among them two have been implemented here. 

The first class of methods consists of replacing the faulty no-stress condition by a better one designed prevent feedback a so-called no-reflexion boundary condition. Such methods have been particularly developed for compressible flows \cite{colonius2004modeling}, where the spurious feedback through pressure takes the form of acoustic waves. Among such methods, the ones based on  {\em characteristics} are the most commonly used. In the incompressible cases, the feedback is to non-local (which can be considered as a limit of the acoustic problem in which the speed of sound is infinite), and the problem is elliptic, so the method of characteristics, which is only relevant to hyperbolic problems, cannot be used. 

A second class of method consists modifying the  equations by introducing a damping term in some regions, usually called {\em sponges} or {\em absorbing layers}, whose purpose is to suppress problematic  features of the flow (here convectively growing perturbations) before they reach the boundaries of the domain. Several ways to modify the equations are possible; in some implementations this is done by artificially increasing the viscosity, while in other this is done by adding a restoring term forcing the flow to relax towards a specified flow (such as a uniform flow or an expected wake profile). In the present work a method of this kind has been implemented, referred to as {\em transverse sponge} (TS),
in which a relaxing term is applied only on the transverse (y) component of the velocity. Accordingly, the Navier-Stokes equation is modified as follows :

\be{NSsponge}
\frac{\partial \bu}{\partial t} + ( \nabla \bu ) \cdot \bu = - \nabla p + Re^{-1} \Delta \bu 
- \underbrace{\gamma_{TS} H_1\left( \frac{x-L_f}{L_t} \right) \left[ \begin{array}{cc} 0 & 0 \\ 0 &1 \end{array} \right] \cdot {\bf u}}_{\mbox{ Transverse sponge}}
\ee

Here $\gamma_{TS}$ is the rate of damping, $L_f$ is the location of the starting of the sponge, $L_t$ is the length of a transition region in which the damping is progressively applied, and $H_1$ is a smooth transition function (whose precise expression does not matter much) verifying $H_1(x)=0$ for $x<0$, $H_1(x)=1$ for $x>1$.
Physically, this modifying term corresponds to assuming a very anisotropic porous medium resisting only to transverse flow. The advantage of this method is that it allows to relax towards a quasi-parallel wake flow without assuming any prescribed profile. Another advantage of the method is that it may be used both for time-stepping nonlinear simulations and for global linear stability.

A third class of methods, applicable only to linear problems, consists of prolongating the solutions as functions of the spatial coordinates $x,y$ considered as complex variables, and to design a mapping between physical coordinates and numerical coordinates $X,Y$ to turn exponentially growing or oscillating terms into damped terms. This is the idea of the {\em perfectly matched layer} (PML) method, popular in electromagnetic and acoustic problems. A variant,  known as the {\em Complex mapping} (CM) method has been introduced in \cite{fabre2019acoustic,sierra2020efficient}. This method maps the axial coordinate $x$ as follows :
\be{CM}
x = X + i \gamma_{CM} H_1\left( \frac{X-L_f}{L_t} \right)  (X-L_f)
\ee 
where $\gamma_s$ is the amplitude, $H_1$ and the subsequent $\bar{H}_1$ are again transition functions allowing to apply the CM from $X>L_f$ with a transition region of width $L_t$. In practice, implementing this method simply consists of modifying all $x$-derivatives as follows:
\be{CM}
\frac{\partial}{\partial x}  \longrightarrow \frac{1}{1 + i \gamma_{CM} \bar{H}_1\left( \frac{X-L_f}{L_t} \right)} \frac{\partial}{\partial X}.
\ee 
As shown in \cite{sierra2020efficient}, the efficiency of this method requires to select $\gamma_{CM}$ with a sign opposite to the real part of the expected eigenvalues $\omega$.

\subsubsection{Results}

\begin{figure}
$$
\includegraphics[width=.48\linewidth]{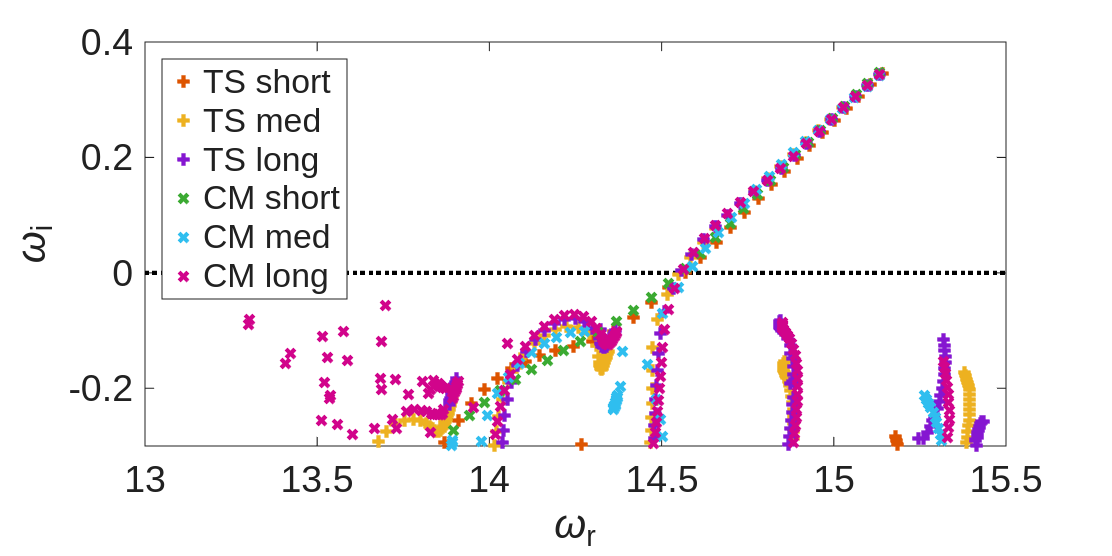}
\includegraphics[width=.48\linewidth]{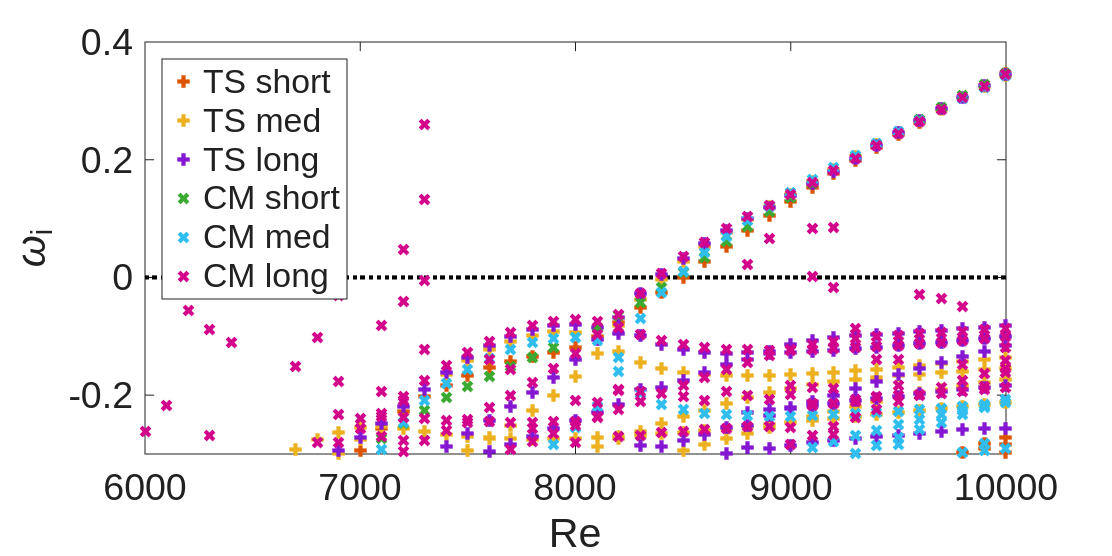}
$$
\caption{Linear spectra for the flow over a NACA0012, using filtering with either a Transverse sponge (TS) or a complex mapping (CM), with several sets of parameters (see text)}
\label{fig:Naca12_Spectra_Filtering}
\end{figure}

In this work, both the TS and the CM method have been implemented. 
Figure \ref{fig:Naca12_Spectra_Filtering} display spectra obtained in this way, with several sets of parameters. The "short" TS-CM are applied quite close to the wing, with $L_f =2$, $L_t=1$, using a rather short numerical domain with $L_x= 5$. The "med" TS-CM are applied farther, with $L_f =4$, $L_t=4$, using a longer numerical domain with $L_x= 10$. Finally the "long" TS-CM were computed with $L_f =9$, $L_t=9$, using a longer numerical domain with $L_x= 40$. The respective amplitudes are chosen as $\gamma_{TS} = 6$ and $\gamma_{CM} = -0.3$, respectively. The figure shows that both methods lead to prediction of a single dominant eigenvalue, hence removing the spurious arc-branch part of the spectrum. Moreover, all sets of parameters lead to very similar results, at least in the most significant half-plane $\omega_i \geq 0$ and predict the onset of unsteadiness for $Re = 8500$.

When scrutinizing the results of figure \ref{fig:Naca12_Spectra_Filtering}, one can observe that the CM results display the smallest sensibility to the parameters. More precisely, the "short" and "medium" CM display the best agreement in the vicinity of the thresholds, which is found by interpolation at $Re_c \approx 8460$ and $8470$, respectively. The "long" CM gives less good results, as it yields a threshold below 8400 and most importantly it leads to the prediction of spurious unstable eigenvalues. Among the results obtained with TS, one observes that the "short" TS leads to results which are closest to the ones obtained with CM, with a threshold for $Re_c \approx 8490$, while the "medium" and "long" TS leads to onset around 8400 or below. 
An interesting conclusion is that the filtering methods perform better when applied rather close to the wing, namely, here, applying the filtering at only 2 cords behind the wing, and truncating the domain to only 5 cords. This goes again against the first expectation that one should apply this filtering rather far from the wing.

\begin{figure}
\includegraphics[width=.48\linewidth]{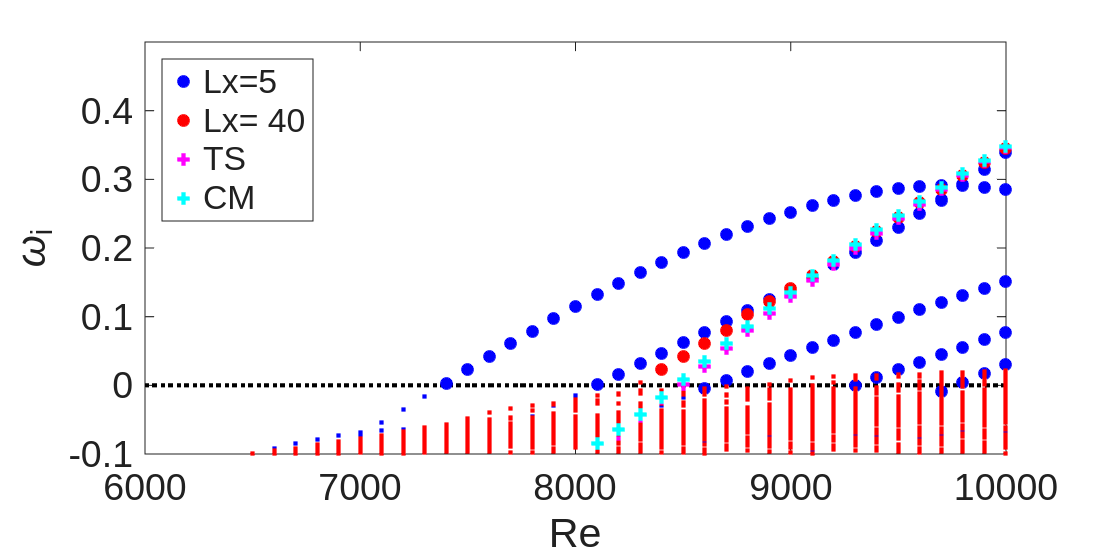}
\includegraphics[width=.48\linewidth]{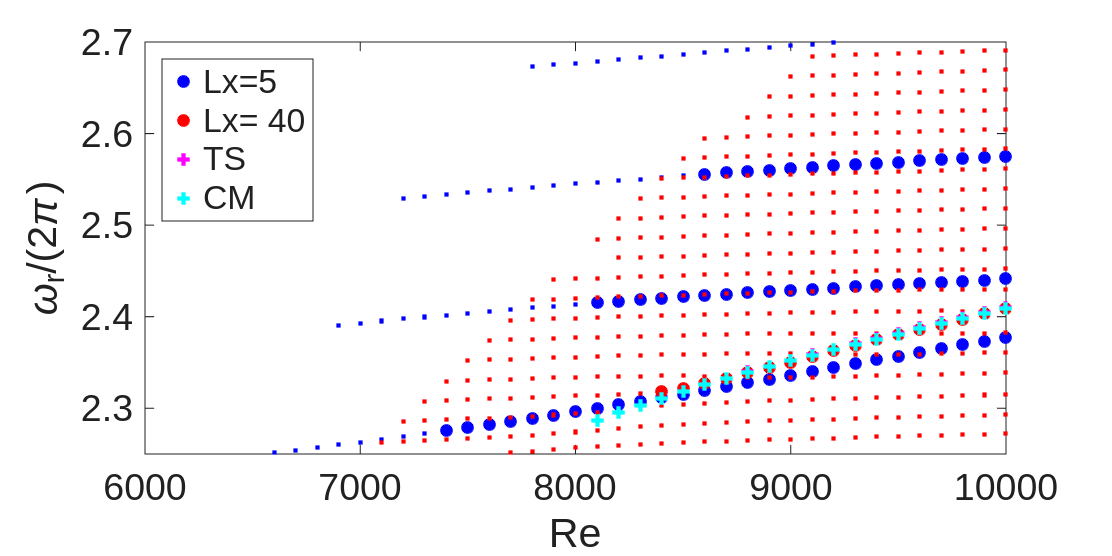}
\caption{Amplification rate $(a)$ and oscillation rate $(b)$ as function of $Re$, with truncated domains of length $L_x= 5$ and $L_x=20$, and with filtering methods. The small dots with same colors as the symbols indicate the damped modes with $\omega_i<0$ (or the weakly amplified mode with $\omega_i<0.025$ for $L_x = 20$).}
\label{fig:Naca12_Branches}
\end{figure}

Figure \ref{fig:Naca12_Branches} compares the results obtained with filtering (here retaining only the "short TS" and "short CM") with those obtained on truncated domains presented in the previous section. Here only results for $L_x=5$ and $L_x=20$ are displayed. As already observed, the results on a truncated domain lead to numerous unstable branches (5 for $L_x= 5$ and about 30 for $L_x=20$) while the results using filtering allow to isolate a single branch. Away from the threshold (for $Re > 9000$), one of the branches obtained on a truncated domain follows well the results obtained with filtering, but departures are observed as $Re$ is decreased and one approaches the threshold. Interestingly, for $L_x= 5$, the branch which fits better with the filtered results is not the first one to emerge but the second one.

\begin{figure}
$$
\includegraphics[width=.8\linewidth]{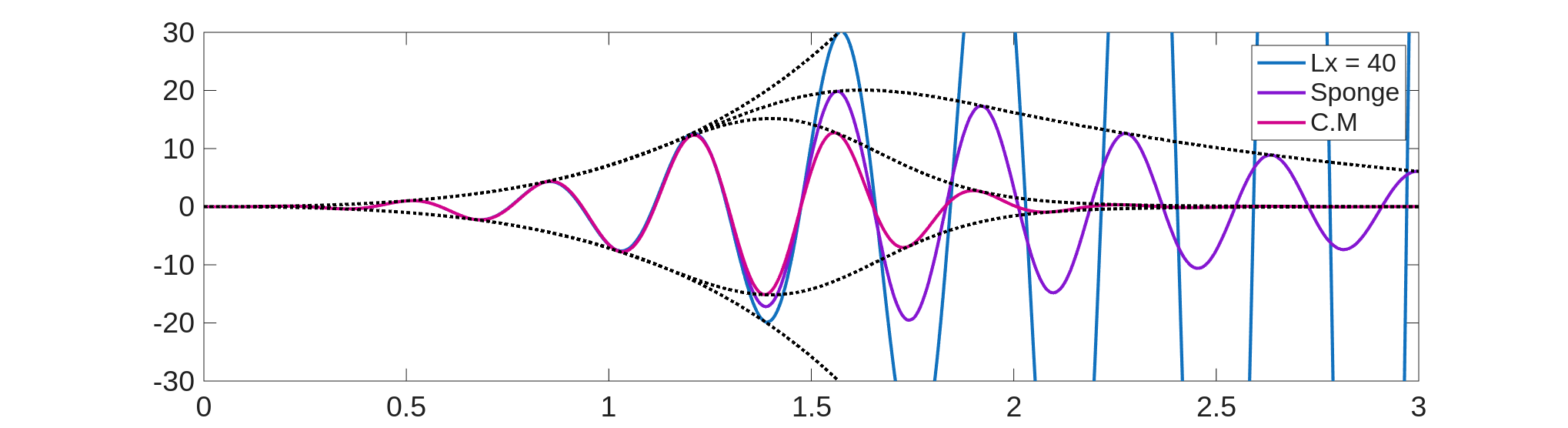}
$$
\caption{Eigenmode with downstream filtering. Here the filtering parameters are $\gamma_{TS}=5,L_s=1,L_t=1$ for transverse sponge (TS) and $\gamma_{TS}=-0.3,L_s=1,L_t=1$ for complex mapping (CM).}
\label{fig:Naca12_Eigenmode8400_Filtering}
\end{figure}

To further demonstrate how the filtering methods operate, figure \ref{fig:Naca12_Eigenmode8400_Filtering}
represents the structure of the leading eigenmode for $Re = 8500$ when computed on the largest domain $L_x= 40$ and using filtering. The result on a truncated domain displays the exponential spatial growth of the perturbation, which was shown to lead to levels of order $10^5$ in figure \ref{fig:Naca12_Eigenmode8400}. The modes computed with filtering initially display the same exponential growth, which quickly turns into an exponential decay as one fully enters the filtering region. It remarkable that the methods correctly predict the eigenvalue and the eigenmode in the most significant region, even with applying the filtering quite close to the wing (here only one wing span downstream).

\section{The NACA0012 profile at small incidence}

\begin{figure}
\includegraphics[width=.48\linewidth]{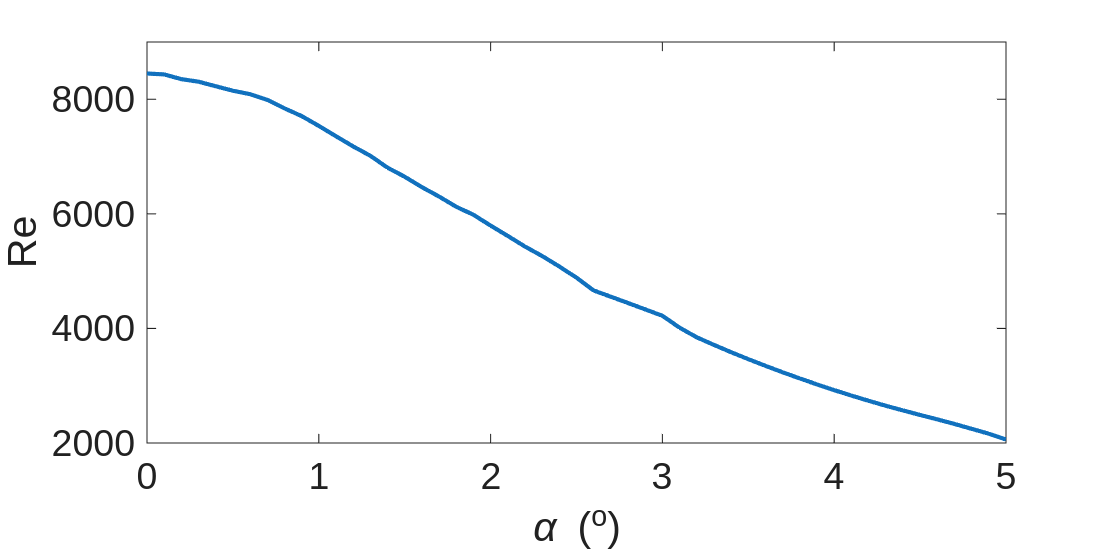}
\includegraphics[width=.48\linewidth]{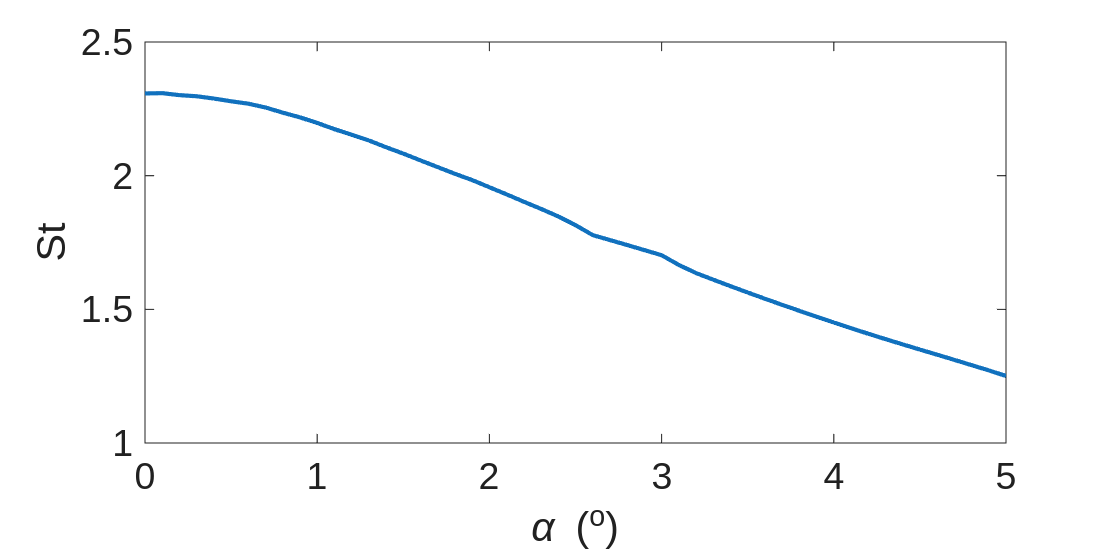}
\caption{Parametric study of the wake of a NACA0012 profile at small incidences:
critical Reynolds $Re_c$ ($a$) and Strouhal number $St_c =  \omega_{r,c}/2\pi$ at onset ($b$) as function of incidence $\alpha$.}
\label{fig:Naca12_Incidence}
\end{figure}

Having designed and validated methods which allow to accurately compute the bifurcation threshold without being affected by spurious effects coming from boundary conditions, we are now able to turn back to the problem of the wake of the NACA0012 profile in the range of low incidence angles $\alpha \in [0^o - 5^o]$, in order to provide a full mapping of the thresholds in this range.  Figure \ref{fig:Naca12_Incidence}$(a)$ displays the neutral curve (here computed using TS) over this range. For $\alpha = 0$ the neutral curve starts at $Re = 8500$, in accordance with the previous section. For $\alpha = 5^o$, which was the lowest value of incidence considered in both \cite{victoria2022stability} and \cite{gupta2023two}, the threshold if found at a value close to $Re = 2000$,  in accordance with both these references. 

As a complement, figure \ref{fig:Naca12_Incidence}$(b)$ documents the oscillation rates at threshold, here represented as $\omega_r/(2 \pi)$ to match with the classical definition of the Strouhal number based on chord length. Here again, the value $St = 1.25$ for $\alpha = 5^o$ is consistent with results from \cite{victoria2022stability} and \cite{gupta2023two}, and the Strouhal continuously rises as the incidence is lowered, towards $St = 2.35$ for the zero-incidence case.

\section{Summary and discussion}

\subsection{Summary}

In the present paper, the problem of the transition in the wake of slender bodies has been thoroughly investigated, using global stability analysis along with other techniques, in order to explain a number of puzzling features identified 
in past studies, which have yet prevented to properly predict the bifurcation thresholds for such bodies. These puzzling features include $(a)$ a strong dependence of stability properties on the domain size, $(b)$ the presence of arc-branch modes in spectra of the linear stability operator, and $(c)$ the absence of a dominating mode clearly linked to absolute instability in a recirculation region, following the generally accepted scenario.

The case of a flat plate of zero-thickness thickness has first be considered, showing that the global spectra are indeed dominated by a large number of arc-branch modes which become unstable almost simultaneously, leading in time-stepping simulations to chaotic fluctuations of the wake in a region which remains confined to the vicinity of the outlet plane. These results are in apparent contradiction with the fact that the flow contains no region of absolute instability.

The nature of these arc-branch modes has then explained through a model combining the convective instability of the jet, well described by local analysis, and a nonlocal instantaneous feedback through a non-local pressure component induced at the outlet boundary as a artefact of the no-stress condition. In combination with a receptivity mechanism occurring at the trailing edge of the plate, this model provides an accurate description of these spurious mode as well as a clear explanation of their occurence. 

The attention was then placed on the wake of the widely used NACA0012 profile. Focussing first on the case of zero 
 incidence, global stability calculations on truncated domain allowed show that the difficulties encountered in this case are due to the interplay between spurious arc-branch modes an a physically relevant eigenmode emerging from the recirculation region. It was eventually shown that the arc-branch modes can be suppressed using specifically designed filtering techniques, including a transverse sponge and a complex mapping technique. Interestingly, and unlike what could be expected, such methods are shown to perform better if the filtering is applied at rather short distances from the wing, of the order of 2 cord spans. These methods eventually give access to the prediction of transition in the very idealized (and hardly approchable in practice) case of a slender body "in an infinite 2D domain", eventually allowing to provide a full mapping of the instability threshold in the range of small incidences.

\subsection{Concluding remarks}

To conclude this paper,   a few implications and perspectives of the present work will be evoked.
At first, through a thorough analysis of possible feedback from boundary conditions and the design of filtering methods allowing to get rid of such spurious effects, the present paper opens a secure route to conduct parametric stability analysis of open shear flow in strongly convective configurations, including wakes, jets, shear layers. This route has already permitted to fill a gap in the literature  in the case of the NACA12 profile at small incidence, one of the most widely used wing model in numerical studies, but for which no mapping of instability properties had been published in the range of small incidences.  Likely, it can be expected that the present ideas will allow to complete parametric stability analyses of many related flows in range of parameters (typically when the Reynolds reaches a few thousands)  where not much has been published, very likely due to the encountering of the difficulties discussed here. 

Along with opening a route to investigate wake and jets in the ideal case of "an infinite domain", the difficulties identified here precisely question the significance of this idealized case. Indeed, in any realistic situation, including real engineering applications and wind-tunnel experiments,  a moving body or a jet never occurs "in an infinite domain". There are necessarily other ingredients, including for instance other distant bodies, walls, an inhomogeneous environment. In the case of wind tunnel experiments, there is also necessarily an outlet which imposes some constraints on the flow, but then the boundary conditions to be applied at the outlet are not controlled, contrarily to numerical simulations. If, as evidenced here, changing the length of the numerical domain from 20 to 40 makes the critical Reynolds number vary by more than 20\%, then a comparable effect can be expected from any of the real-life complications listed above. Moreover, in realistic situations, in the range of Reynolds number of a few thousands, unsteadiness is more often observed to be intermittent, or modulated in time as in the numerical simulations for a flat plate presented at the beginning of \S 2. Owing to all these issues, 
the linear stability framework might well just not be the right tool for such problems, and other approaches, such as ones based on the resolvent, might be more suited to describe the dynamics.

Finally, another major contribution of the present paper is the highlighting of a possible long-range feedback occurring through the pressure in incompressible flows. Feedback through pressure is well known to occur in the compressible case and is commonly explained in terms of acoustic waves propagating upwards. On the other hand, to the knowledge of the author, a similar feedback in the strictly incompressible case, which becomes instantaneous and nonlocal, has not been so well documented in incompressible flows.  In the present study, the nonlocal feedback was introduced to model a spurious effect originating from the boundary conditions in numerical simulations. However, there are many other situations where a nonlocal feedback could be invoked to explain the onset of self-sustained oscillations. For instance, for a jet impacting on a distant plate, numerous unstable modes are observed  with a discretization of frequencies clearly depending on the distance between the nozzle and the plate. In the weakly compressible case, this can be explained by the interaction of a downstream propagating vorticity wave and in upstream propagating acoustic wave \citep{ausin2023mode}, but the latter does not exist any more in the zero-mach limit. The present framework may prove to be useful in such situation, by deriving models for the strictly incompressible problem rather than thing to work out the singular limit of a compressible problem.

\section*{Acknowedgements}
This work has been a long-lasting effort, requiring more than 3 years to tackle a problem looking simple but proving to be very challenging. Over this long period, the reflection has been constantly alimented by fruitful discussions with numerous colleagues, which turned out to be decisive at some stages. The author particularly wishes to thank a number of collegues, starting with {\sc Olivier Marquet} and {\sc Diogo Ferreira Sabino} with whom the difficulties of the problem were initially identified. Later on, regular discussion with {\sc Flavio Giannetti} and {\sc Javier Sierra Ausin} have been particularly stimulating. Special thanks to {\sc Javier} for developing and kindly sharing the implicit time-stepper! {\sc Vincenzo Citro}, {\sc Paolo Luchini} and {\sc Lutz Lesshafft} are also thanked for more punctual but fruitful discussions.

\section*{Declaration of Interests.} The authors report no conflict of interest.

\noindent This is an Open Access article written under the terms of the CC-BY 4.0 licence (https://creativecommons.org/licenses/by/4.0/).

\bibliographystyle{jfm}
\bibliography{biblio}

\end{document}